\documentclass[nofootinbib,floatfix,aps,prd,twocolumn,superscriptaddress]{revtex4-1}
\usepackage{graphicx}
\usepackage{textcomp}
\usepackage{amsmath}
\usepackage{amssymb}
\usepackage{amsthm}
\usepackage{float}
\usepackage{braket}
\usepackage{fancyhdr}
\usepackage{slashed}
\usepackage{mathrsfs}
\usepackage{dsfont}
\usepackage{multirow} 
\usepackage{color}
\usepackage{xcolor}
\usepackage[subfigure]{graphfig}

\newcommand{\be}{\begin{equation}}
\newcommand{\ee}{\end{equation}}
\newcommand{\ben}{\begin{equation*}}
\newcommand{\een}{\end{equation*}}
\newcommand{\bea}{\begin{eqnarray}}
\newcommand{\eea}{\begin{eqnarray}}

\newcommand{\oct}{O_h^D}
\newcommand{\Mheavy}{a094m358}
\newcommand{\Mlight}{a094m278}
\newcommand{\lattMom}{\left(2\pi/L\right)^{-1}ap_z}

\begin{document}
\title{\Large Distillation at High-Momentum}

\newcommand*{\JLAB}{Thomas Jefferson National Accelerator Facility, Newport News, VA 23606, USA}\affiliation{\JLAB}
\newcommand*{\WM}{Physics Department, William and Mary, Williamsburg, Virginia 23187, USA}\affiliation{\WM}
\author{Colin Egerer}\affiliation{\WM}
\author{Robert G. Edwards}\affiliation{\JLAB}
\author{Kostas Orginos}\affiliation{\JLAB}\affiliation{\WM}
\author{David G. Richards}\affiliation{\JLAB}

\collaboration{On behalf of the \textit{HadStruc Collaboration}}

\begin{abstract}
  Extraction of hadronic observables at finite-momenta from Lattice QCD
  (LQCD) is constrained by the well-known signal-to-noise problems
  afflicting all such LQCD calculations. Traditional quark smearing
  algorithms are commonly used tools to improve the statistical quality
  of hadronic $n$-point functions, provided operator momenta are
  small. The momentum smearing algorithm of Bali et al.\ extends the
  range of momenta that are cleanly accessible, and has facilitated
  countless novel lattice calculations. Momentum smearing has, however,
  not been explicitly demonstrated within the framework of
  distillation. In this work we extend the momentum-smearing idea, by
  exploring a few modifications to the distillation framework. Together
  with enhanced time slice sampling and expanded operator bases
  engendered by distillation, we find ground-state nucleon energies can
  be extracted reliably for $\left|\vec{p}\right|\lesssim3\text{ GeV}$
  and matrix elements featuring a large momentum dependence can be
  resolved.
\end{abstract}
\maketitle

\section{Introduction}
Lattice field theory is now a thoroughly well-established scheme to
quantitatively study strongly-interacting theories, such as Quantum
Chromodynamics (QCD), from first-principles. With the exception of the
lightest pseudoscalar mesons at rest, lattice QCD (LQCD) calculations
of the spectrum and properties of hadrons are afflicted by
exponentially worsening signal-to-noise ratios as the Euclidean time
extent between operators grows. It is thus a key demand of lattice
calculations that the hadron of interest saturate correlation functions at
as short a Euclidean time separation as possible.  Key to satisfying
this demand is identifying an operator whose overlap with the hadron
of interest is maximized relative to those with other states:
$\bra{0}\hat{\mathcal{O}}\left(\vec{p}\right)\ket{h\left(\vec{p}\right)}\gg\bra{0}\hat{\mathcal{O}}\left(\vec{p}\right)\ket{h'\left(\vec{p}\right)}$.

The most widely used means of accomplishing this is through quark
spatial smearing schemes, such as Wuppertal~\cite{Gusken:1989qx} or
Jacobi~\cite{Allton:1993wc} smearing, which act as low-energy filters
of hadronic correlation functions, leading to a more rapid relaxation
to low-energy eigenmodes. It is thus standard practice to compute
hadronic observables where at least one interpolating operator of an
N-point function possesses a non-trivial spatial extent. However, as
pointed out in ref.~\cite{Bali:2016lva}, spatial smearing of hadronic
operators is less than optimal and even detrimental for all but
interpolators projected to zero momentum. The authors proposed a
remedy, now known as momentum smearing, that involves the introduction
of appropriately tuned phase factors onto the underlying gauge links,
prior to the subsequent spatial smearing of the quark fields.  In
effect, a tunable momentum space distribution is constructed by
creating an oscillatory spatial profile. The remarkable effectiveness
of this procedure was established in~\cite{Bali:2016lva}, wherein the
pion and nucleon energies were reliably extracted up to $\sim2\text{
  GeV}$ and $\sim3\text{ GeV}$, respectively, and the dispersion
relations reasonably satisfied.

This robust momentum-smearing technique is now ubiquitous in lattice
studies that demand a wide range of momenta, such as the mapping of
nucleon electromagnetic form factors (FFs)~\cite{Kallidonis:2018cas},
generalized FFs~\cite{Bali:2018zgl}, and semi-leptonic decay FFs
needed to quantify elements of the Cabibbo-Kobayashi-Maskawa
matrix~\cite{Bahr:2019eom,Bazavov:2019aom}. Perhaps the greatest
usage has been seen in LQCD calculations of matrix elements of certain non-local
space-like-separated operators, which when computed over a range of momenta can be
related to various light-cone distributions fundamental
to hadron structure. Such matrix elements, analyzed
in the context of Large Momentum Effective Field Theory
(LaMET)~\cite{Ji:2013dva,Ji:2014gla}, have proven useful in
understanding the (un)polarized partonic content of the pion and
nucleon~\cite{Izubuchi:2019lyk,Lin:2018qky,Alexandrou:2018eet,Chen:2018xof,Alexandrou:2018pbm,Fan:2018dxu}.
Quark bilinears can be related via coordinate space factorization
schemes to lightcone distribution
amplitudes~\cite{Bali:2019dqc,Bali:2018qat,Bali:2018spj}, and to quark
parton distribution functions
(PDFs)~\cite{Joo:2019bzr,Joo:2019jct,Sufian:2019bol} within the
pseudo-PDF framework~\cite{Radyushkin:2017cyf}. Whilst the ``Lattice
Cross Sections'' approach~\cite{Ma:2014jla,Ma:2017pxb} generalizes
this paradigm to spatially separated gauge-invariant
current-current matrix elements, recently employed in~\cite{Sufian:2019bol,Sufian:2020vzb}
to determine the valence quark content of the pion.

Although momentum smearing, in concert with Wuppertal or Jacobi
smearing, does indeed enhance the overlap of the interpolating
operators onto the lowest lying states in the spectrum, there are
additional challenges that it does little to ameliorate.  Firstly, energy
eigenstates contributing to a correlator become dense as the
spatial momentum of the correlators increases.  Secondly, the reduced
lattice symmetries for correlators at non-zero spatial momentum,
together with the contribution of two- and higher-particle states,
further increases the density of the higher energies.
Distillation~\cite{Peardon:2009gh} when employed with an extended
basis of operators that it facilitates, provides a powerful means of
addressing these issues, as well as permitting a better sampling of a
gauge configuration through explicit momentum projections
performed at both source and sink in a two-point correlation
function. The use of the variational method within a given lattice
symmetry channel, using an extended basis of operators implemented
through distillation, has
proven essential in mapping the low-lying baryon spectrum of
QCD~\cite{Edwards:2011jj,Dudek:2012ag} and exotic
hadrons~\cite{Dudek:2009qf,Dudek:2010wm,Dudek:2011tt,Liu:2012ze}, as
well as exploring the glueball content in the isoscalar sector of
QCD~\cite{Dudek:2013yja}. Recently, the power of this approach has been
demonstrated in the calculation of the various nucleon
isovector charges\cite{Egerer:2018xgu}.
Calculational programs employing
distillation have generically limited the spatial momenta to
within the shell $\left| a_s \vec{p}\right|^2\lesssim4 (2 \pi/L_s)^2$,
where $a_s$ is the spatial lattice spacing, and $L_s$ is the number of
time slices in the spatial directions.  Here the resultant correlation
functions have sufficient momentum-space overlap that the distillation
framework does not necessitate modifications. The goal of
this work is to supplement distillation with a realization of
momentum smearing, thereby increasing the range of hadron momenta
accessible, and in so doing demonstrate the efficacy of this approach both
for the nucleon energies at higher spatial momenta and for the
nucleon charges derived at these high momenta.

The remainder of this paper is organized as follows. We proceed
in Section~\ref{sec:Dist} with a brief summary of the distillation
framework, and the modifications needed to incorporate momentum smearing
within that framework. In Section~\ref{sec:proofOfPrinciple}, we
describe its computational implementation, and then proceed to a
comparison of the nucleon energies with and without momentum smearing on a
lattice at the larger of our two pion masses, and
identify an optimal procedure for its implementation.  
In Section~\ref{sec:charges}, we extend the investigation to a lighter
pion mass, and in particular highlight the
efficacy of this approach by determining the renormalized isovector
charges of the nucleon in both stationary and boosted-frames, with and without the
momentum-smearing modifications. In Section~\ref{sec:chargeBehavior} 
we discuss our results for the resultant matrix elements, and their
interpretation in terms of both the expected discretization effects,
and the possible excited-to-ground-state transitions. Concluding remarks are
given in Section~\ref{sec:Conclude}.

\section{Distillation\label{sec:Dist}}
Distillation~\cite{Peardon:2009gh} is a low-rank approximation to a
gauge-covariant smearing kernel, conventionally taken to be the
Jacobi-smearing kernel
$J_{\sigma,n_\sigma}\left(t\right)=\left(1+\frac{\sigma\nabla^2\left(t\right)}{n_\sigma}\right)^{n_\sigma}$~\cite{Allton:1993wc}. The
tunable parameters $\lbrace\sigma,n_\sigma\rbrace$ allow for variable
source ``widths'' and applications, respectively, such that in the
large iteration limit, the kernel approaches that of a
spherically-symmetric Gaussian.
The low-rank approximation is formed by isolating eigenvectors
of the discretized three-dimensional gauge-covariant Laplacian
\[
-\nabla^2 (t)\xi^{(k)}\left(t\right)=\lambda^{(k)}(t)\xi^{(k)}\left(t\right)
\]
and ordering solutions according to the eigenvalue magnitudes
$\lambda^k\left(t\right)$. The outer product of equal-time
eigenvectors defines the distillation smearing kernel
\be
\label{eq:distop}
\Box\left(\vec{x},\vec{y};t\right)_{ab}=\sum_{k=1}^{R_{\mathcal{D}}}\xi_a^{\left(k\right)}\left(\vec{x},t\right)\xi_b^{\left(k\right)\dagger}\left(\vec{y},t\right),
\ee where $R_{\mathcal{D}}$ is the chosen rank of the distillation
space and color indices $a,b$ are made explicit. Correlation functions
formed by Wick-contracting quark fields smeared via~\eqref{eq:distop}
can be factorized into distinct reusable components, the
\textit{elementals} and the \textit{perambulators}. The elementals
\begin{equation}
  \Phi_{\alpha\beta\gamma}^{\left(i,j,k\right)}\left(t\right)=\epsilon^{abc}\left(\mathcal{D}_1\xi^{\left(i\right)}\right)^a\left(\mathcal{D}_2\xi^{\left(j\right)}\right)^b\left(\mathcal{D}_3\xi^{\left(k\right)}\right)^c\left(t\right)S_{\alpha\beta\gamma},\label{eq:elementals}
\end{equation}
shown here for the case of baryons, encode the operator construction,
where $\mathcal{D}_i$ are covariant derivatives, and
$S_{\alpha\beta\gamma}$ are subduction coefficients encoding how an interpolator
with Dirac indices $\lbrace\alpha,\beta,\gamma\rbrace$
constructed in the continuum will mix across irreducible
representations (irreps) of a hypercubic lattice and its associated
little groups.  The perambulators \be
\tau_{\alpha\beta}^{\left(l,k\right)}\left(t',t\right)=\xi^{\left(l\right)\dagger}\left(t'\right)M^{-1}_{\alpha\beta}\left(t',t\right)\xi^{\left(k\right)}\left(t\right)
\ee encode the propagation of the quarks between elements of the
distillation space, where $M$ is the Dirac operator.  It is this
factorization of the quark propagation from the construction of the
interpolating operators that enables the computationally efficient
implementation of the variational method with an extended basis of
operators.

\subsection{Momentum Smeared Distillation}
Distillation is quite costly initially both in
computational storage and the construction of its components.
Moreover, the rank $R_\mathcal{D}$ is expected to scale with the lattice spatial
volume in order to maintain the same resolution in correlation
functions on different ensembles~\cite{Peardon:2009gh}. This is
particularly significant for the construction of the correlation
functions, where the
needed Wick contractions for meson and baryon two-point functions scale as
$R_\mathcal{D}^3$ and $R_\mathcal{D}^4$, respectively. Thus an
implementation of momentum smearing within distillation must seek to
minimize the number of additional distillation vectors included in
the basis, and in particular avoid the use of a distinct eigenvector
basis for each momentum of the correlation functions.

\begin{figure}[h!]
  \centering
  \includegraphics[width=\linewidth]{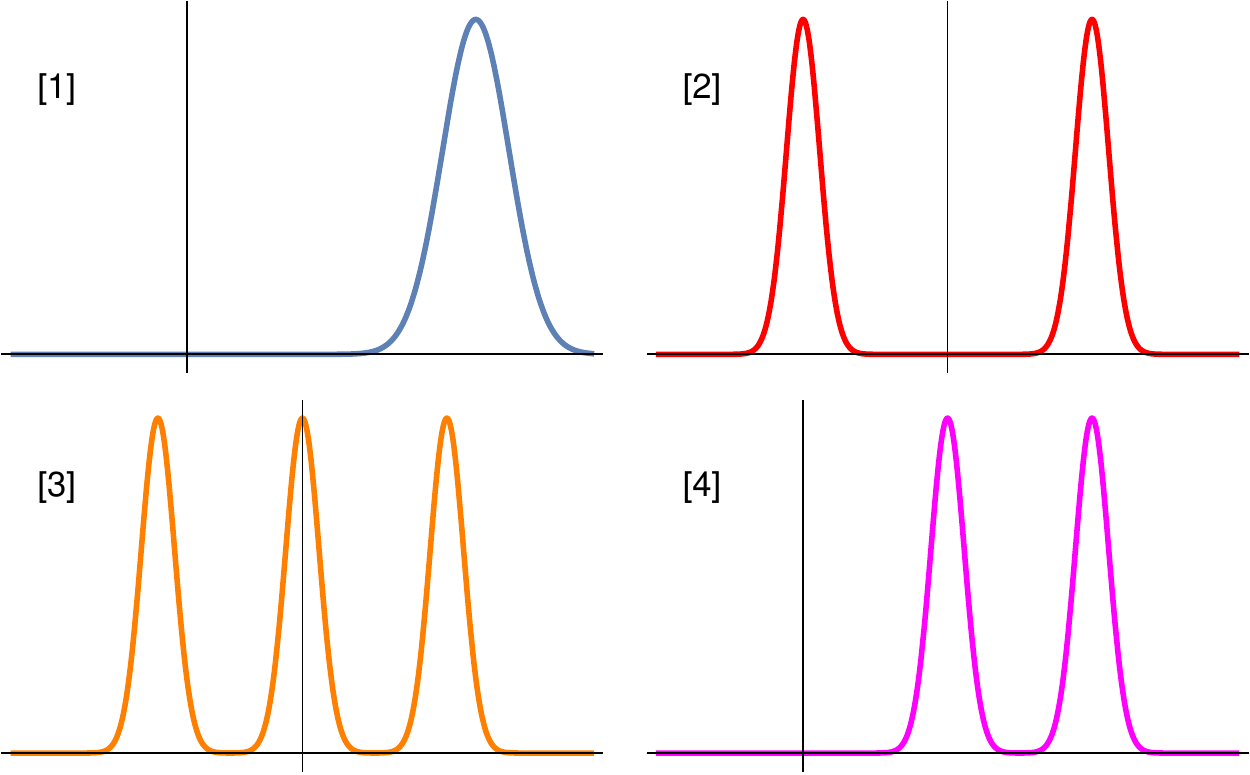}
  \caption{Qualitative momentum space overlaps following modification
    of a computed eigenvector basis. Panels 2-4 expressly violate
    translation invariance, but would dramatically reduce
    computational cost were translational symmetry
    preserved.\label{fig:momsmear_types}}
\end{figure}

With such a scenario in mind, one might consider modifying a set of eigenvectors according to:
\begin{enumerate}
\item\textit{Single Phase}
  \ben
  \tilde{\xi}_a^{\left(k\right)}\left(\vec{z},t\right)=e^{i\vec{\zeta}\cdot\vec{z}}\xi_a^{\left(k\right)}\left(\vec{z},t\right)
  \een
\item\textit{Opposing Phases}
  \ben
  \tilde{\xi}_a^{\left(k\right)}\left(\vec{z},t\right)=2\cos\left(\vec{\zeta}\cdot\vec{z}\right)\xi_a^{\left(k\right)}\left(\vec{z},t\right)
  \een
\item\textit{Identity and Opposing Phases}
  \ben
  \tilde{\xi}_a^{\left(k\right)}\left(\vec{z},t\right)=\left[1+2\cos\left(\vec{\zeta}\cdot\vec{z}\right)\right]\xi_a^{\left(k\right)}\left(\vec{z},t\right)
  \een
\item\textit{Multiple Unidirectional Phases}
  \ben
  \tilde{\xi}_a^{\left(k\right)}\left(\vec{z},t\right)=\left[e^{i\vec{\zeta_1}\cdot\vec{z}}+e^{i\vec{\zeta_2}\cdot\vec{z}}\right]_{\zeta_1\neq\zeta_2}\xi_a^{\left(k\right)}\left(\vec{z},t\right),
  \een
\end{enumerate}
such that overlaps for several, potentially opposing, hadron momenta
could be simultaneously improved. A schematic qualitative picture of
these candidate implementations is
depicted in Fig.~\ref{fig:momsmear_types}.

An important requirement of any modification of distillation is the
preservation of translational invariance, since that is essential for the
projection to states to definite momentum.  It is straightforward to
show that the perambulators with the type-1 modification are indeed
invariant under the translation of the phase through $\vec{x} \rightarrow \vec{x} + \vec{d}$:
\begin{align*}
  \tilde{\tau}_{\mu\nu}^{ij}\left(t',t\right)=\xi^{\left(i\right)\dagger}\left(\vec{x},t'\right)&e^{-i\vec{\zeta}\cdot\left(\vec{x}+\vec{d}\right)}M^{-1}_{\mu\nu}\left(\vec{x},t';\vec{y},t\right)
  \\&\times
  e^{i\vec{\zeta}\cdot\left(\vec{y}+\vec{d}\right)}\xi^{\left(j\right)}\left(\vec{y},t\right)
  \\ =\xi^{\left(i\right)\dagger}\left(\vec{x},t'\right)&e^{-i\vec{\zeta}\cdot\vec{x}}M^{-1}_{\mu\nu}\left(\vec{x},t';\vec{y},t\right)e^{i\vec{\zeta}\cdot\vec{y}}\xi^{\left(j\right)}\left(\vec{y},t\right).
\end{align*}
Such translation invariance fails for the other implementations of
momentum smearing, as we show below for phasing of Type 4:
\begin{align*}
  \tilde{\tau}_{\mu\nu}^{ij}&\left(t',t\right)=\xi^{\left(i\right)\dagger}\left(\vec{x},t'\right)\lbrace
  e^{-i\vec{\zeta_2}\cdot\left(\vec{x}+\vec{d}\right)}+e^{-i\vec{\zeta_1}\cdot\left(\vec{x}+\vec{d}\right)}\rbrace
  \\ &\times M^{-1}_{\mu\nu}\left(\vec{x},t';\vec{y},t\right)\lbrace
  e^{i\vec{\zeta_1}\cdot\left(\vec{y}+\vec{d}\right)}+e^{i\vec{\zeta_2}\cdot\left(\vec{y}+\vec{d}\right)}\rbrace\xi^{\left(j\right)}\left(\vec{y},t\right)
  \\ &\quad\quad\thickspace=\xi^{\left(i\right)\dagger}\left(\vec{x},t'\right)e^{-i\vec{\zeta}_2\cdot\vec{x}}e^{i\left(\vec{\zeta}_1-\vec{\zeta}_2\right)\cdot\vec{d}}M^{-1}_{\mu\nu}\left(\vec{x},t';\vec{y},t\right)
  \\ &\times
  e^{i\vec{\zeta}_1\cdot\vec{y}}\xi^{\left(j\right)}\left(\vec{y},t\right)+\lbrace\vec{\zeta}_1\leftrightarrow\vec{\zeta}_2\rbrace+\mathcal{T.I.}
\end{align*}
where we find a 
combination of translationally invariant ($\mathcal{T.I.}$) and
variant pieces for $\vec{\zeta}_1\neq\vec{\zeta}_2$.
Thus in the remainder of this paper, we consider only phasing of type
1, and refer to the modified eigenvector basis as ``phased''.

The momentum smearing scheme of ref.~\cite{Bali:2016lva} reweights
gauge fields $U_\mu\left[x\right]$ in a boost direction $z_\mu$ with
weight $\zeta=\tfrac{2\pi}{L}\mathfrak{r}$ according to \be
\tilde{U}_\mu\left[x\right]=e^{i\frac{2\pi}{L}\mathfrak{r}
  z_\mu}U_\mu\left[x\right] \ee prior to quark source creation, where $\mathfrak{r}\in\mathbb{R}$.
As phases are applied to the underlying gauge configurations prior to
determination of the eigenvectors, the configurations can safely be
smeared with unallowed lattice momenta as highlighted in~\cite{Bali:2016lva}.
Thus it is sufficient to modify the previously computed eigenvectors,
limiting phases to allowed lattice momenta. In particular we consider the phase factors
\begin{align}
  &\vec{\zeta}=\frac{2\pi}{L}\hat{z}\label{eq:onePhase}, \\
  &\vec{\zeta}= 2\cdot\frac{2\pi}{L}\hat{z}\label{eq:twoPhase},
\end{align}
corresponding to one and two units of the allowed lattice momenta.
We remark that phases applied in the $-\hat{z}$-direction improve
momentum space overlaps for $ap_z<0$ but are not presented herein for
brevity.

\section{Demonstration of Efficacy\label{sec:proofOfPrinciple}}
We employ two isotropic clover ensembles, with $2 \oplus 1$ flavors,
of extent $32^3 \times 64$, an inverse coupling $\beta = 6.3$,
corresponding to a lattice spacing $a\simeq0.094~{\rm fm}$, and with
pion masses of 358 and $278~{\rm MeV}$, respectively. These are cataloged
in Table~\ref{tab:ensembles}; further details of the ensembles are
contained in ref.~\cite{Yoon:2016dij,Yoon:2016jzj}.  To first
establish the feasibility of our candidate implementation, we employ
the ensemble at the heavier pion mass, herein denoted by $\Mheavy$.
The figure of merit we use is to extract the ground-state nucleon
dispersion relation for as large a range of momentum as possible.
\begin{table}
  \begin{center}
    \setlength\abovecaptionskip{-1pt}
    \setlength{\belowcaptionskip}{-10pt}
    \begin{tabular}{c|c|c|c|c|c|c}\hline\hline
      ID  & $a$ (fm)  &$m_\pi$ (MeV)  & $L^3\times N_t$ & $N_{\rm cfg}$ & $N_{\rm srcs}$ & $R_{\mathcal{D}}$ \\
      \hline
      $a094m358$ & $0.094(1)$ & 358(3) & $32^3\times 64$ & 100 & 4 & 64 \\
      $a094m278$ & $0.094(1)$ & 278(4) & $32^3\times 64$ & 259 & 4 & 64 \\
      \hline\hline
    \end{tabular}
  \end{center}
  \caption{Lattice ensembles utilized throughout this work. The number of distillation eigenvectors $R_{\mathcal{D}}$ and distinct source positions $N_{\rm srcs}$ per configuration are also indicated.\label{tab:ensembles}}
\end{table}

Calculations were performed for four distinct (randomized) source
temporal origins on 100 configurations of the $\Mheavy$
ensemble, with each configuration separated by $10$ HMC
trajectories; this small number of configurations was found sufficient
to quantitatively demonstrate the effectiveness of distillation for
the nucleon energies and dispersion relation. We employed $R_{\mathcal{D}} = 64$ eigenvectors,
where the gauge fields in the Laplacian
were smoothed via $10$ iterations of stout
smearing~\cite{Morningstar:2003gk} with smearing parameter
$\rho_{ij}=0.08$ and $\rho_{\mu4}=\rho_{4\mu}=0$.

\subsection{Interpolator Construction}
The regularization of QCD through lattice discretization
explicitly breaks continuum rotational symmetry, and consequently
baryons at rest are now cataloged according to the double-cover
irreps of the octahedral group $\oct$.
Thus mass eigenstates once cataloged by $J^P$ must now be isolated
according to their patterns of subduction across the finite number of
irreps $\Lambda$ of $\oct$. The construction of the nucleon operators
follows the procedure introduced in refs.~\cite{Edwards:2011jj,Dudek:2012ag}, which
we summarize now, and are expressed in terms of the baryon elementals
introduced in Eq.~\ref{eq:elementals}.
These operators are projections onto the lattice irreps of discretized
continuum-like operators, which we classify according to
the spectroscopic notation $N^{(2S+1)}L_{\mathcal{P}}J^P$, where $S$
represents the Dirac spin, $L$ the angular momentum introduced via
derivatives, $\mathcal{P}$ the permutational symmetry of such
derivatives, and $J^P$ the total angular momentum and parity of the
nucleon interpolator $N$.

To best capture the ground-state $J^P=\tfrac{1}{2}^+$ nucleon at rest,
which trivially subduces into the $G_{1g}$ irrep of $\oct$, we use a
basis of non-relativistic interpolators~\cite{Edwards:2011jj,Dudek:2012ag}:
\begin{align}
  \mathcal{B}_{\vec{p}=\vec{0}}=\lbrace N^2S_S\tfrac{1}{2}^+&,N^2S_M\tfrac{1}{2}^+,N^2S_S'\tfrac{1}{2}^+,N^2P_A\tfrac{1}{2}^+,\nonumber\\
  &N^2P_M\tfrac{1}{2}^+,N^4P_M\tfrac{1}{2}^+,N^4D_M\tfrac{1}{2}^+\rbrace
  \label{eq:restBasis}
\end{align}
that admit a flexible description of the radial/orbital nucleon
structure - we note $N^2P_M\tfrac{1}{2}^+$ and $N^4P_M\tfrac{1}{2}^+$
are of hybrid construction.

Projection of the lattice interpolating fields to non-zero spatial
momenta ($\vec{p}\neq\vec{0}$) further breaks the $\oct$ symmetry group
to little groups dependent on the
$\textbf{*}\left(\vec{p}\right)$\cite{Moore:2005dw}, and furthermore
mixes states of different parities.  Here we consider only boosts
along a spatial axis, which are especially important for PDF
calculations in the LaMET and pseudo-PDF frameworks.  In this case,
the little group is the order-$16$ dicyclic group or
$\text{Dic}_4$.  The framework for the construction of the operators,
specialized to the case of mesons,
is given in ref.~\cite{Thomas:2011rh}.
The genesis is the classification of
operators of definite helicity, and therefore we extend our basis
both to include those of higher spins, and of negative parity, which
are then subduced to the little group. In particular, our basis is
extended as follows, based on the study of the
nucleon spectrum and the dominant operators in ref.~\cite{Dudek:2012ag}:\footnote{
  Note $N^2S_S'\tfrac{1}{2}^+$ is removed from our interpolator basis}
\begin{align}
  \mathcal{B}_{\vec{p}\neq\vec{0}}=\lbrace&N^2S_S\tfrac{1}{2}^+,N^2S_M\tfrac{1}{2}^+,N^2P_A\tfrac{1}{2}^+,N^2P_M\tfrac{1}{2}^+,\nonumber\\
  &N^4P_M\tfrac{1}{2}^+,N^4D_M\tfrac{1}{2}^+,N^4S_M\tfrac{3}{2}^+,N^2D_S\tfrac{5}{2}^+,\nonumber\\
  &N^2P_M\tfrac{1}{2}^-,N^4P_M\tfrac{1}{2}^-,N^2P_M\tfrac{3}{2}^-,N^4P_M\tfrac{3}{2}^-,\nonumber\\
  &N^4P_M\tfrac{5}{2}^-,N^2D_S\tfrac{3}{2}^+,N^4D_M\tfrac{3}{2}^+,N^2D_M\tfrac{3}{2}^+\rbrace.
  \label{eq:inflightBasis}
\end{align}
We emphasize that the density of the (discrete) energy spectrum for
the nucleon is expected to be considerably greater for states in
motion compared with those at rest for the following reasons.
Firstly, as the spatial momentum is increased the separation between
the energies of a given state is compressed.
Secondly, through the reduced symmetries, even in the continuum,
that enables more states to contribute within a given symmetry channel.

\subsection{Variational Analysis}
The factorization of a correlation function intrinsic to distillation
facilitates the use of an extended basis of interpolators at source
and sink, without re-computation of quark propagators as in standard
smearing schemes. We are then able to perform a 
variational analysis in the nucleon $G_{1g}$ channel at rest
(Eq.~\ref{eq:restBasis}), and for all boosted frames in the
$\text{Dic}_4$ little group (Eq.~\ref{eq:inflightBasis}). We start
with a matrix of correlation functions
\begin{equation}
  C_{ij}(T,\vec{p})=\bra{0}\mathcal{O}_i(T,-\vec{p})\mathcal{O}_j^\dagger(0,\vec{p})\ket{0},
\end{equation}
where $\vec{p}$ is the momentum projection, and $\mathcal{O}^\dagger$ selected from some
interpolator basis $\mathcal{B}$; we reiterate that distillation
enables momentum projections at both source and sink time slices, respectively.
The variational method corresponds to solution of a generalized eigenvalue problem (GEVP)
of the form \be C(T,\vec{p})v_{\bf
  n}\left(T,T_0\right)=\lambda_{\bf
  n}\left(T,T_0\right)C(T_0,\vec{p})v_{\bf n}\left(T,T_0\right).
\label{eq:GEVP}
\ee
Optimal operators, in the variational sense, for the energy eigenstates $\ket{\bf n}$
are defined by
$\sum_iv_{\bf n}^i\mathcal{O}_i^\dagger$.
Associated with each eigenvector is a \textit{principal correlator}
$\lambda_{\bf n}\left(T,T_0\right)$.
We will obtain the energy associated with each state
$\ket{\bf n}$ by fitting its
principal correlator according to
\begin{equation}
  \lambda_{\bf
    n}\left(T,T_0\right)=\left(1-A_{\bf n}\right)e^{-E_{\bf
      n}\left(T-T_0\right)}+A_{\bf n}e^{-E_{\bf n}'\left(T-T_0\right)}.
  \label{eq:princorrFit}
\end{equation}
The inclusion of a second exponential serves to
quantify the extent to which a principal correlator is
dominated by a single state, for which any deviation is encapsulated
by the amplitude $A_{\bf n}$ and ``excited'' energy $E_{\bf n}'$.
Further details, and in particular regarding the selection of $t_0$
and the conditions used to enforce orthogonality of eigenvectors
$v_{\bf n}\left(T,T_0\right)$, are contained in
refs.~\cite{Dudek:2010wm,Egerer:2018xgu}.

\begin{figure*}[th!]
  \centering
  \subfigure[]{\includegraphics[width=0.49\linewidth]{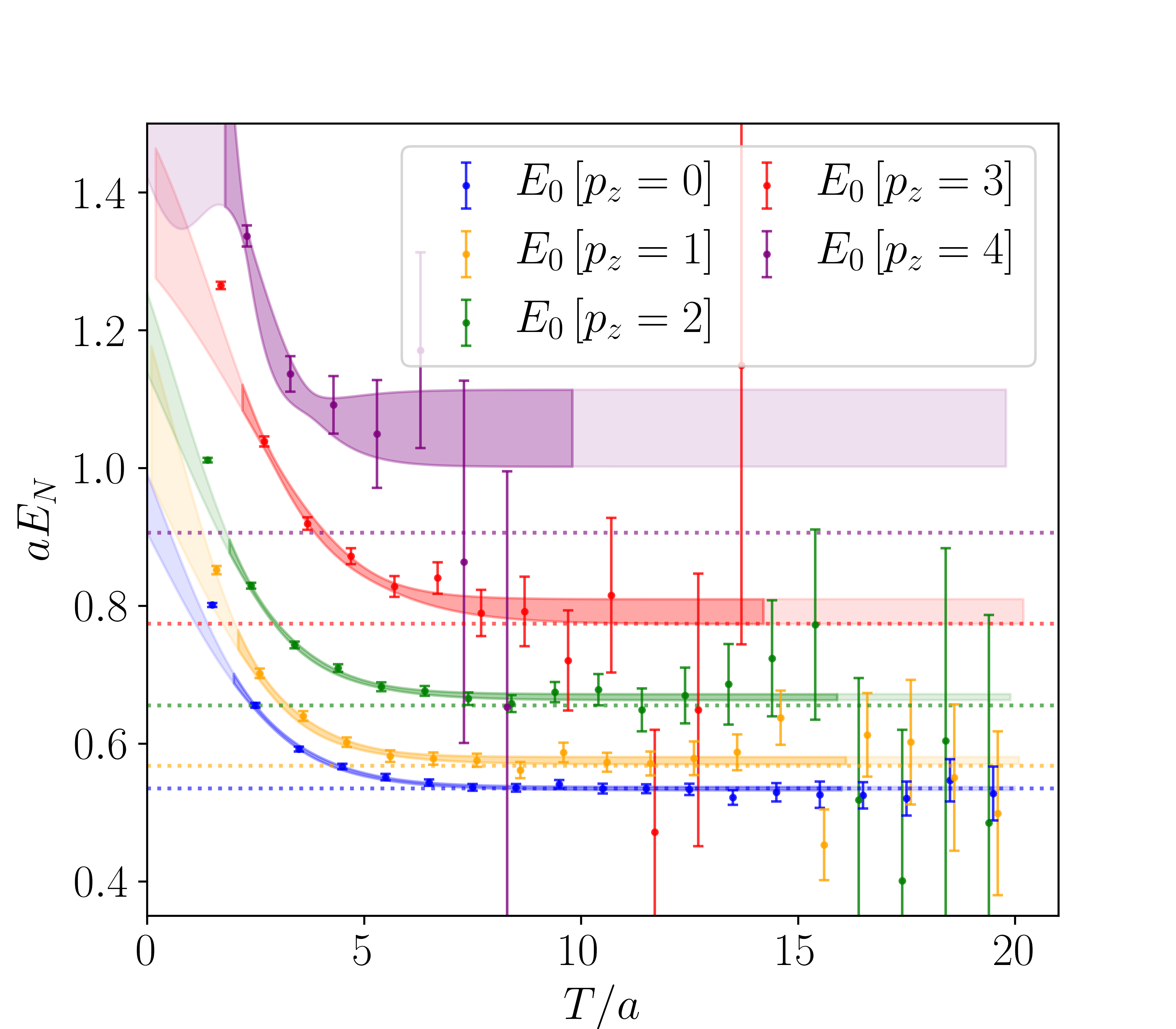}\label{fig:effectiveenergies2350_local}}
  \subfigure[]{\includegraphics[width=0.49\linewidth]{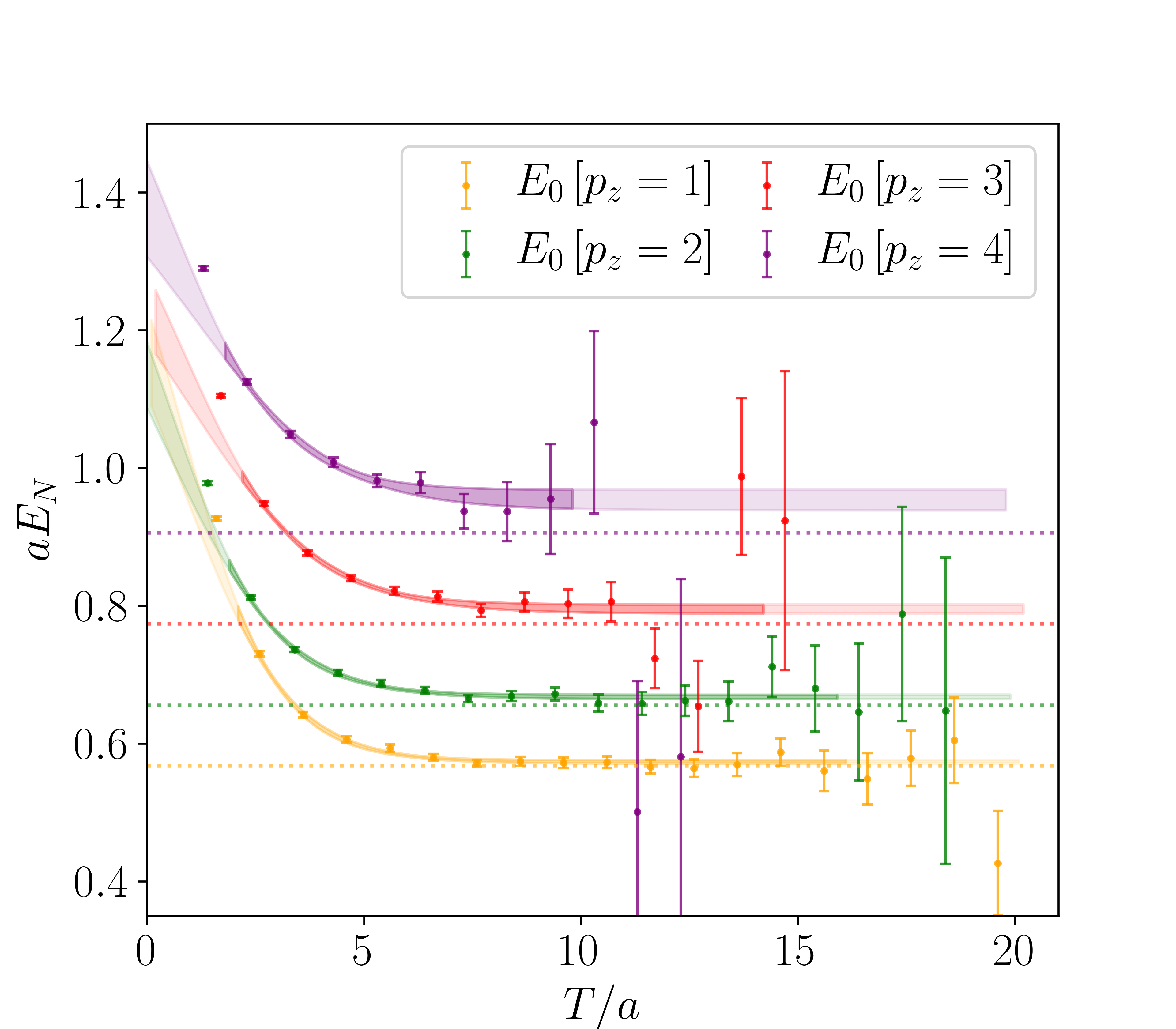}\label{fig:effectiveenergies2350_local_d001_1p00}}
  \caption{The left-hand (a) and right-hand (b) plots show the
    effective energies for the nucleon, obtained on the $\Mheavy$ ensemble, using a single, local
    interpolating operator $N^2S_S\frac{1}{2}^+$, subduced to the
    relevant little group, constructed with
    unphased (a) and phased (b) distillation eigenvectors, respectively.  Data
    are shown for points where the signal-noise ratios are $\geq1.35$
    (a) and $\geq2$ (b), and are shifted for legibility.
    The bands show the two-state fits to the correlators, as
    described in the text, where the dark region indicates data
    included in the fits.  The dashed lines represent the energies expected
    from the continuum dispersion relation using the nucleon mass obtained from the fit
    to the $\vec{p}=0$ correlator.}
\end{figure*}

\begin{figure*}[th!]
  \centering
  \subfigure[]{\includegraphics[width=0.495\linewidth]{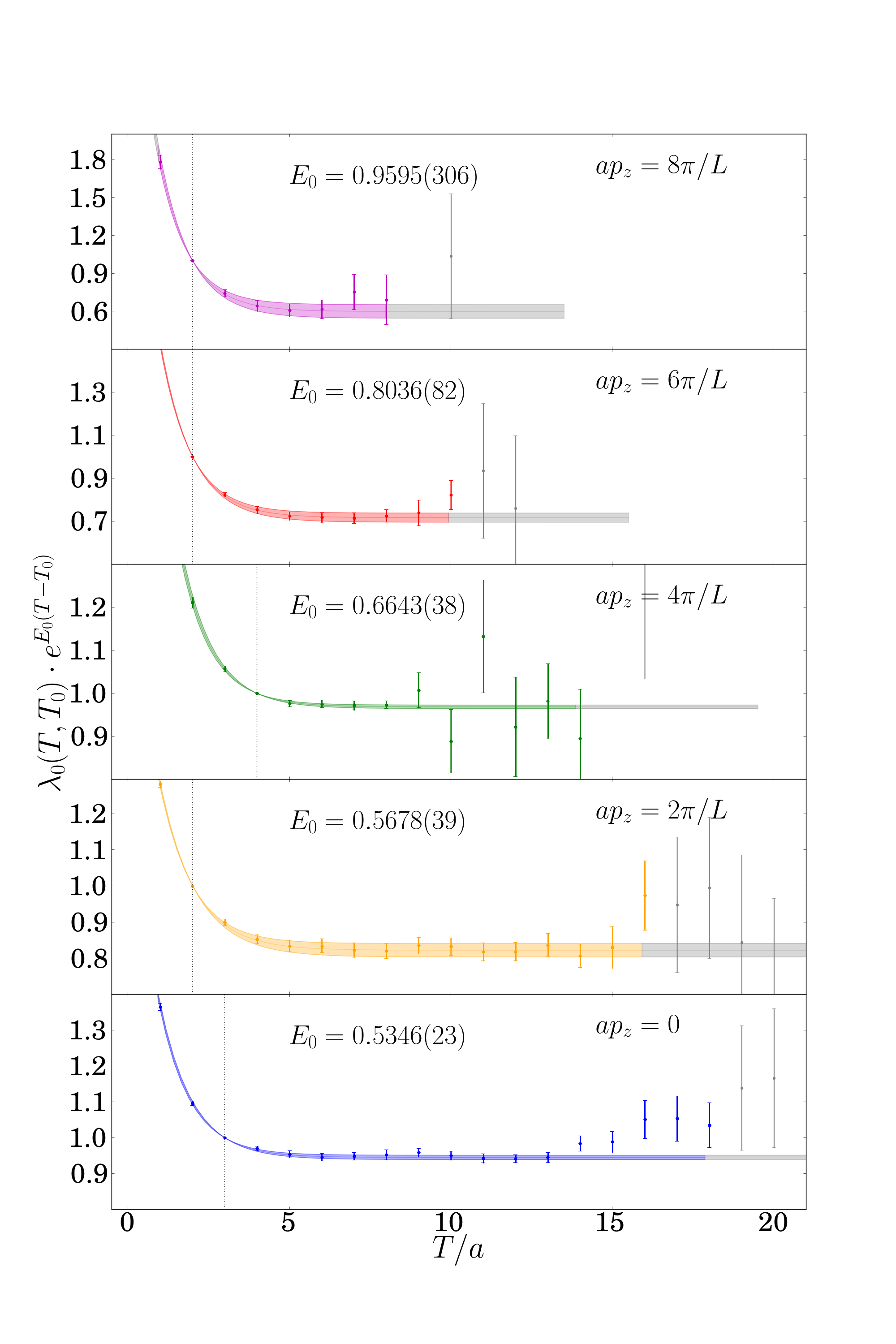}\label{fig:princorrs2350_unphased}}
  \subfigure[]{\includegraphics[width=0.495\linewidth]{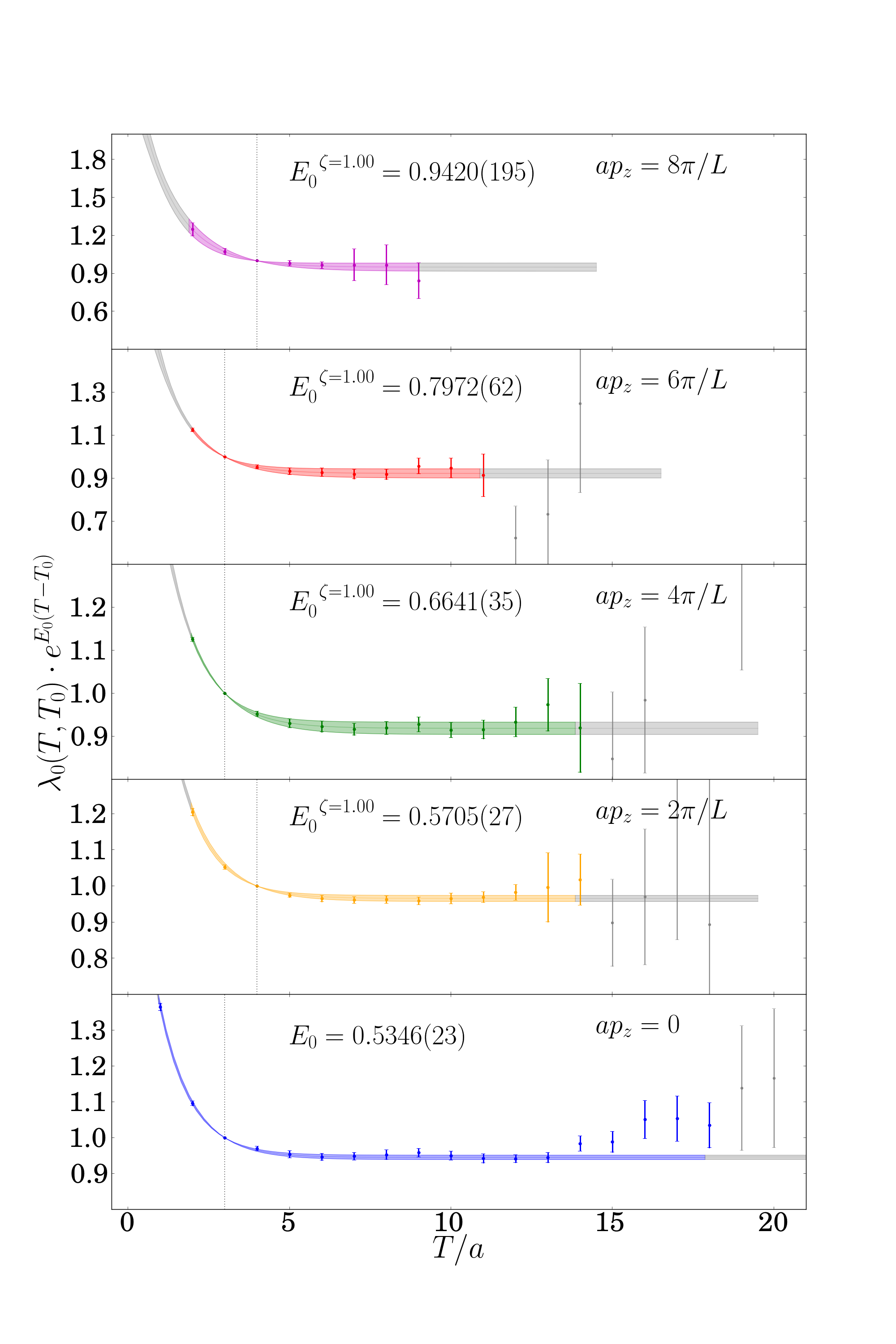}\label{fig:princorrs2350_d001_1p00}}
  \caption{The ground-state nucleon principal correlators for the
    $\Mheavy$ ensemble using a projected interpolator within each momentum channel obtained from the
    $\mathcal{B}_{\vec{p}\neq\vec{0}}$ interpolator basis subduced into
    the relevant little group. The left-hand and right-hand panels are obtained
    from the unphased and phased eigenvectors, with one unit of momentum, respectively.
    The ground-state principal
    correlator for the unphased $\mathcal{B}_{\vec{p}=\vec{0}}$ basis
    is shown for reference (blue).  In each case, data are shown for
    signal-to-noise ratios $\geq2$. The bands show the two-exponential
    fits of Eq.~\ref{eq:princorrFit}, with data excluded from the fits
    in grey. Both the data and fits are shown as $\lambda_0 e^{E_0(T-T_0)}$,
    where $E_0$ is the lowest-lying energy obtained from the
    fit.\label{fig:princorrs2350}}
\end{figure*}

\subsection{Efficacy of Phased Distillation \& Nucleon Dispersions\label{subsec:nucEffEnergies}}
We benchmark the standard distillation implementation, without phasing, by first
computing ground-state nucleon energies using the single, local interpolating operator
$N^2S_S\tfrac{1}{2}^+$, the analog to standard nucleon
interpolators, for $ap_z\le4\left(2\pi/L\right)$.
We fit the two-point functions to the two-exponential form
\begin{equation}
  \label{eq:2statefit}
  C_{\text{fit}}^{2\text{pt}}(T,\vec{p})=e^{-E(\vec{p})T}\left(a+be^{-\Delta
    ET}\right),
\end{equation}
where $\Delta E$ is the gap between the ground and excited-state
energies, and priors are introduced to ensure the positivity of the
overlap parameters $\lbrace a,b\rbrace$. To avoid possible contact
terms arising from the use of the Wilson-clover action, only temporal
separations greater than one are included in the fit.  The data and
the resulting fits are shown in
Figure~\ref{fig:effectiveenergies2350_local}.  For the lowest
momenta $ap_z\le2\left(2\pi/L\right)$, the data exhibit a clear signal
over the large range of $T/a$, and are well described by a two-state fit.
Furthermore, the resulting ground-state energies 
are in excellent agreement with the expectations from the
continuum dispersion relation $E^2=m^2+p^2$. However, for momenta
$ap_z=\{3,4\}\times\left(2\pi/L\right)$, not only does the signal-to-noise
ratio degrade rapidly, but a two-state fit becomes insufficient to
capture the contributions of excited states to the correlator
signal. The latter is seen by the tension between the fit and
correlator for Euclidean separations $T/a\leq5$. Inclusion of
additional states in the functional of~\eqref{eq:2statefit} would
undoubtedly better describe early times in the
$ap_z=\{3,4\}\times\left(2\pi/L\right)$ signals, but the lack of statistically
meaningful signal beyond $T/a\simeq10$ presents a serious
limitation.

Figure~\ref{fig:effectiveenergies2350_local_d001_1p00} features the
$N^2S_S\frac{1}{2}^+$ correlators where the underlying eigenvectors
are phased with one unit of momentum, as in
Eq.~\ref{eq:onePhase}. While there is only a modest improvement in
the statistical precision of large-$T/a$ signal for $ap_z=\{1,2\}\times\left(2\pi/L\right)$,
a dramatic improvement is seen for the
$ap_z=\{3,4\}\times\left(2\pi/L\right)$ signals. The improved statistical
precision with phasing also serves to expose deviations of the
energies from the expectations of the continuum dispersion
relation. These discrepancies could arise from discretization effects,
or from incomplete determination of the ground state correlation
function.  It is this latter possibility that we now try to control
through the use of the variational method.

We performed the variational analysis on the matrix of correlation
functions formed by interpolators in the
$\mathcal{B}_{\vec{p}=\vec{0}}$ (Eq.~\ref{eq:restBasis}) and
$\mathcal{B}_{\vec{p}\neq\vec{0}}$ bases
(Eq.~\ref{eq:inflightBasis}). We first applied the variational method
to the unphased basis to determine the improvement this provides with
respect to the single operator used above.   We
then performed the same analyses with distillation spaces modified
according to~\eqref{eq:onePhase} (one unit of momentum) and
~\eqref{eq:twoPhase} (two units of momentum), over the momentum ranges
$1\leq\lattMom\leq4$ and $4\leq\lattMom\leq8$, respectively.  These
momentum ranges were chosen to emphasize that, although one would
naively expect eigenvectors modified according to~\eqref{eq:onePhase}
to have optimal overlap with momenta $ap_z=3\left(2\pi/L\right)$
and~\eqref{eq:twoPhase} with $ap_z=6\left(2\pi/L\right)$, a broad
coverage in momentum is possible within each modified space, thereby
obviating the need to use many distillation bases each with its
own computational cost.

The principle correlators, together with the two-state fits of
Eq.~\ref{eq:princorrFit}, are shown in the left and right-hand plots
of Figure~\ref{fig:princorrs2350} for the cases of unphased
eigenvectors, and phased eigenvectors with one unit momentum,
respectively.  Compared to the use of phasing with the single
$N^2S_S\frac{1}{2}^+$ interpolator, the gains afforded by a
variational analysis of the phased operator basis appear less dramatic
than use of an unmodified basis. The principal correlators in each
case demonstrate a rather uniform plateau very close to unity,
indicative of single eigenstate dominance. However, the phased
principal correlators are much better determined and lead to more
precise determinations of the ground-state nucleon energies. For
example in the $ap_z=4\left(2\pi/L\right)$ case, the extracted nucleon energy from
the phased principal correlator is $\sim35\%$ more precise than the
unphased equivalent.

\begin{figure}[th!]
  \includegraphics[width=\linewidth]{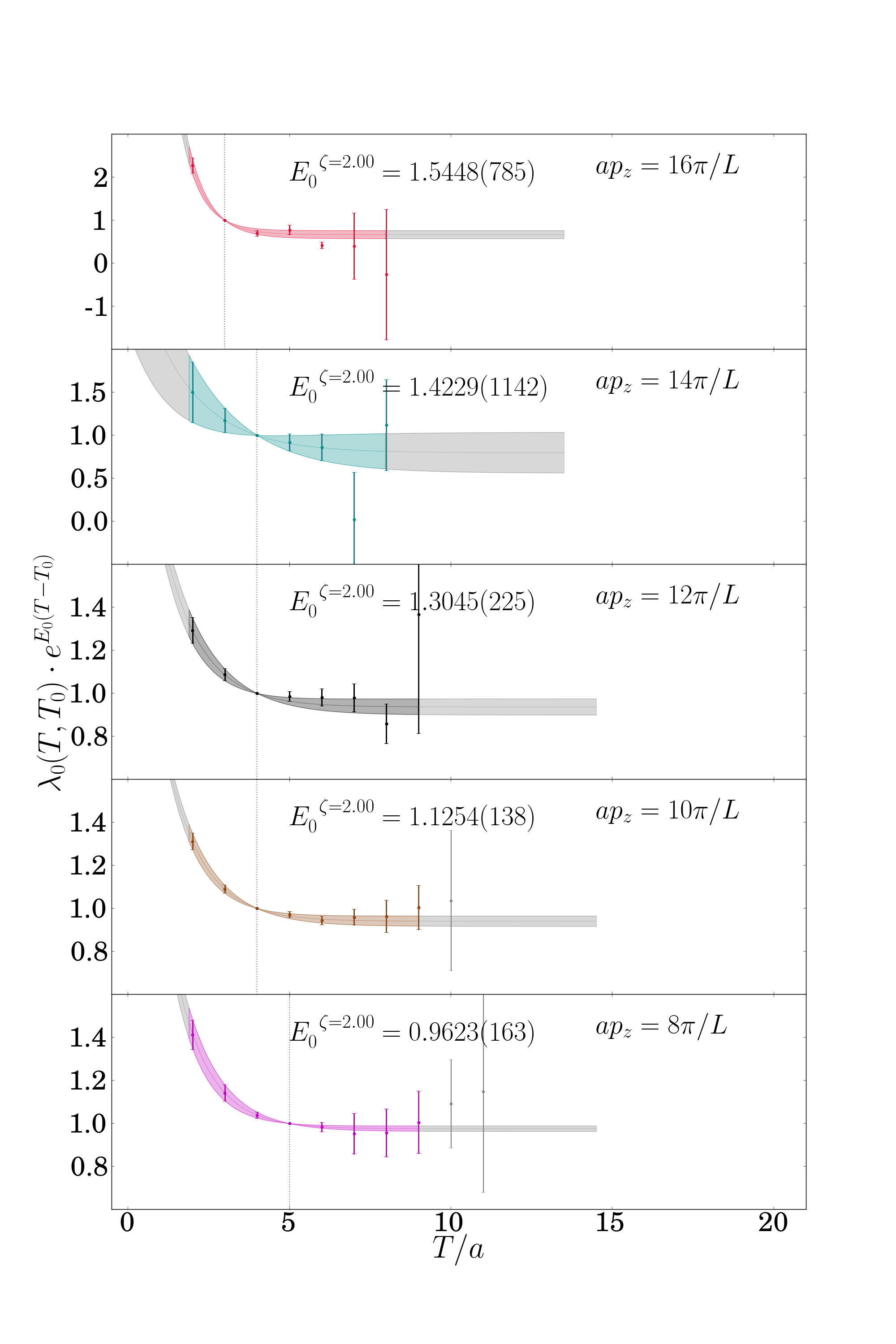}
  \caption{The ground-state nucleon principal correlators for the
    $\Mheavy$ ensemble using a projected interpolator within each momentum channel obtained from
    the $\mathcal{B}_{\vec{p}\neq\vec{0}}$ interpolator basis subduced
    into the relevant little group. The eigenvectors are phased with
    two units of momentum~\eqref{eq:twoPhase}. Principal correlator fits
    ~\eqref{eq:princorrFit} are shown with colored bands, while
    excluded data are in grey. Data is shown for signal-to-noise ratios
    $\geq2$.\label{fig:princorrs2350_d001_2p00}}
\end{figure}

For the highest momenta $4\leq\lattMom\leq8$ a comparison with the
unphased principle correlators is not possible due to expected statistical
fluctuations. We instead show in
Fig.~\ref{fig:princorrs2350_d001_2p00} principal correlators for
$4\leq\lattMom\leq8$, where now the eigenvectors are phased with two
units of allowed lattice momenta~\eqref{eq:twoPhase}.  Though the
principle correlators for the higher excited states could not be
resolved in such highly boosted frames, the resolution of the ground-state nucleon to at least
$ap_z=6\left(2\pi/L\right)$ marks a considerable improvement in the distillation/GEVP
infrastructure for the study of hadron structure.

\begin{figure}[h]
  \centering
  \includegraphics[width=\linewidth]{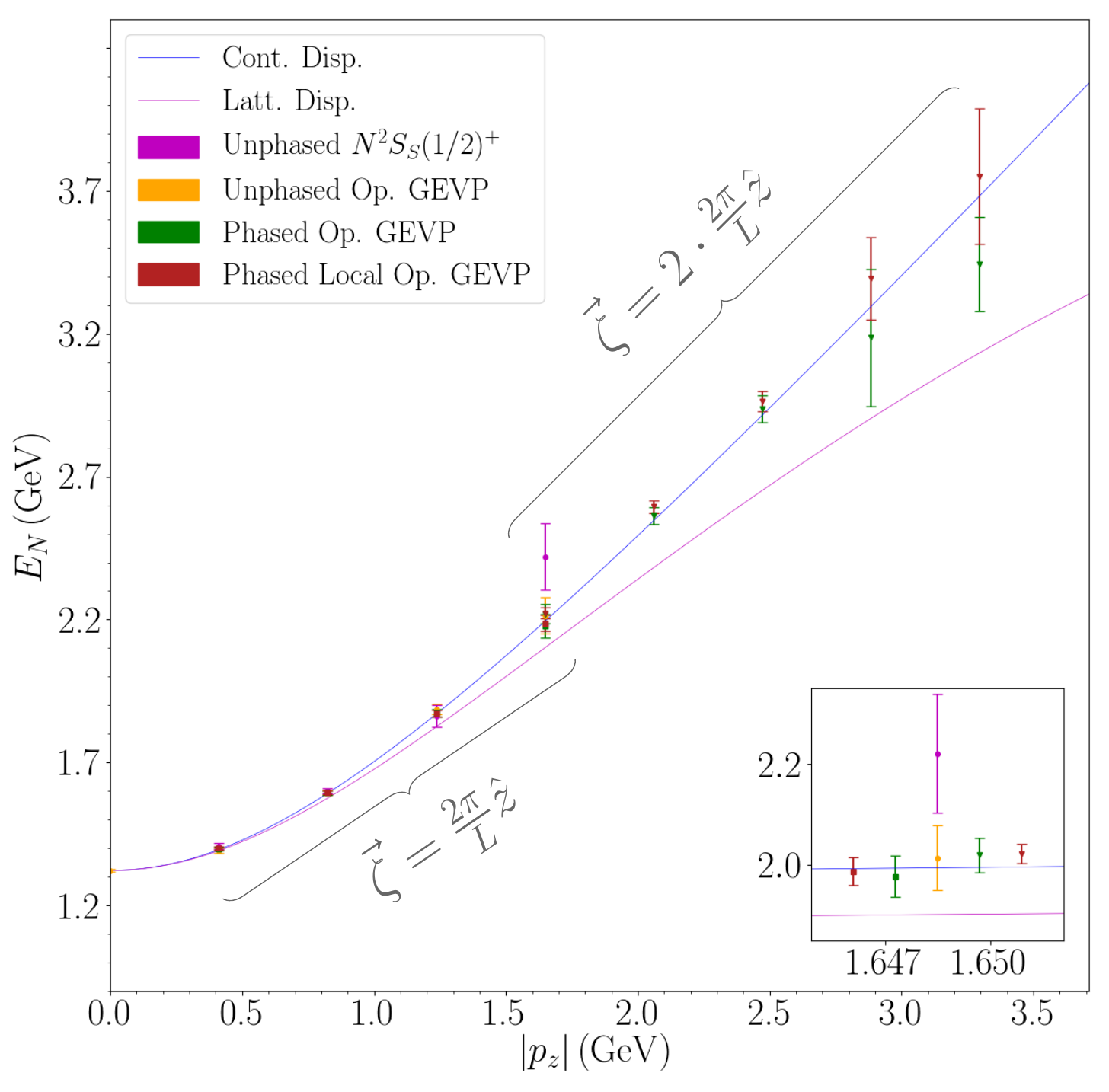}
  \caption{The ground-state nucleon dispersion relation for the
    $\Mheavy$ ensemble, together with expectations from the continuum
    dispersion relation (blue), and free lattice scalar dispersion
    relation (purple). Energies without the use of phasing are shown
    in magenta for a single, $N^2S_S\tfrac{1}{2}^+$ operator, and
    orange for the variational analysis using the bases
    $\mathcal{B}_{\vec{p}=\vec{0}},\mathcal{B}_{\vec{p}\neq\vec{0}}$.
    The energies obtained by applying the variational method on the phased
    $\mathcal{B}_{\vec{p}=\vec{0}},\mathcal{B}_{\vec{p}\neq\vec{0}}$
    bases are shown in green, and for a basis of purely local
    operators in red.  The squares and triangles denote the
    $\vec{\zeta}=\frac{2\pi}{L}\hat{z}$ and
    $\vec{\zeta}=2\cdot\frac{2\pi}{L}\hat{z}$ phasing, respectively.  The
    ground-state nucleon energies for momentum
    $ap_z=4\left(2\pi/L\right)$ are shown in the inset plot,
    shifted for legibility.\label{fig:m0p2350_dispersion_plots}}
\end{figure}

The results for our variational analyses of the unphased and phased
bases for different momenta are summarized in
Fig.~\ref{fig:m0p2350_dispersion_plots}, where we plot the extracted
nucleon energies, together with expectations from both the continuum
dispersion relation, and the lattice dispersion relation for a free
scalar particle.  It is evident, even with the use of an extended
operator basis and the correspondingly improved isolation of
the ground state, distillation without phasing is unable to cleanly resolve the
ground-state nucleon energy for $ap_z=4\left(2\pi/L\right)$, where
the signal is dominated by noise whether the single
or variationally optimized operator is used.

Energies from the low-momentum phasing~\eqref{eq:onePhase} were
found to be consistent with those determined from the unphased GEVP,
but are of substantially higher statistical quality. Most encouraging
is that we are now able to map the ground-state nucleon dispersion
relation up to $p_z \simeq 3\text{ GeV}$ using the 
$\vec{\zeta}=2\cdot\frac{2\pi}{L}\hat{z}$ phased distillation space, even within
the limited statistics. Moreover, significant
uncertainty in the nucleon energies accrues only for
the highest momenta $ap_z=\{7,8\}\times\left(2\pi/L\right)$, where for
discretization effects are considerable.

Confidence in our extracted nucleon energies is bolstered by a
separate variational analysis of an extended operator basis containing
only the spatially-local interpolators, in particular the
$N^2S_S\tfrac{1}{2}^+$ and seven explicitly relativistic
interpolators. These results are shown in red of
Fig.~\ref{fig:m0p2350_dispersion_plots}, and are again consistent with
the (un)phased determinations when using the
$\mathcal{B}_{\vec{p}\neq\vec{0}}$ operator basis. The slightly higher
values for the nucleon energies at large momenta are not surprising,
as the purely local operator basis did not include negative-parity
operators nor those of continuum spin $J>\tfrac{3}{2}$, certainly
contaminating the true ground-state nucleon signal. Nonetheless, a
consistent determination of the nucleon dispersion relation when using
two distinct operator bases validates the union of distillation with
momentum smearing, and in particular confirms that the addition of
phase factors does not spoil the group theory required to construct
our interpolating operators.

\section{Matrix Elements at High Momentum\label{sec:charges}}
Hadron structure calculations within lattice QCD proceed through
calculation of matrix elements between hadrons of interest,
implemented through the calculation of three-point, or higher,
correlation functions.  As emphasized in the introduction, many of the
key measures of hadron structure, such as the parton distribution
functions computed in the LaMET, pseudo-PDF or lattice-cross-section
frameworks, require that the resulting three-point functions be
computed for hadrons at as large a momentum, or over as large a range
of momentum, as possible in order to have the best control over
systematic uncertainties in their approaches.  Thus the remainder of
this paper is devoted the addressing this issue through the
calculation of the nucleon isovector charges, in the forward direction, both for the nucleon at
rest and for the nucleon in a moving frame of increasing boosts.

For our study of the nucleon charges, we use an ensemble at a somewhat
lighter pion mass, which we denote by $\Mlight$, for which the relevant
isovector current renormalization constants have been
computed~\cite{Yoon:2016jzj}; details of the ensemble are contained in
Table~\ref{tab:ensembles}.  At the lower values of momentum
($ap_z=\{0,1\}\times\left(2\pi/L\right)$), we use
the vanilla form of distillation, without phasing.
As we demonstrate below, at high momentum, where phasing is essential,
we use two units of phasing, as implemented in
Eq.~\ref{eq:twoPhase}. For $ap_z=4\left(2\pi/L\right)$, we compare
our results both with and without phasing as a consistency check of
the method.

\subsection{Nucleon Effective Energies}
\begin{figure}[b]
  \centering
  \includegraphics[width=\linewidth]{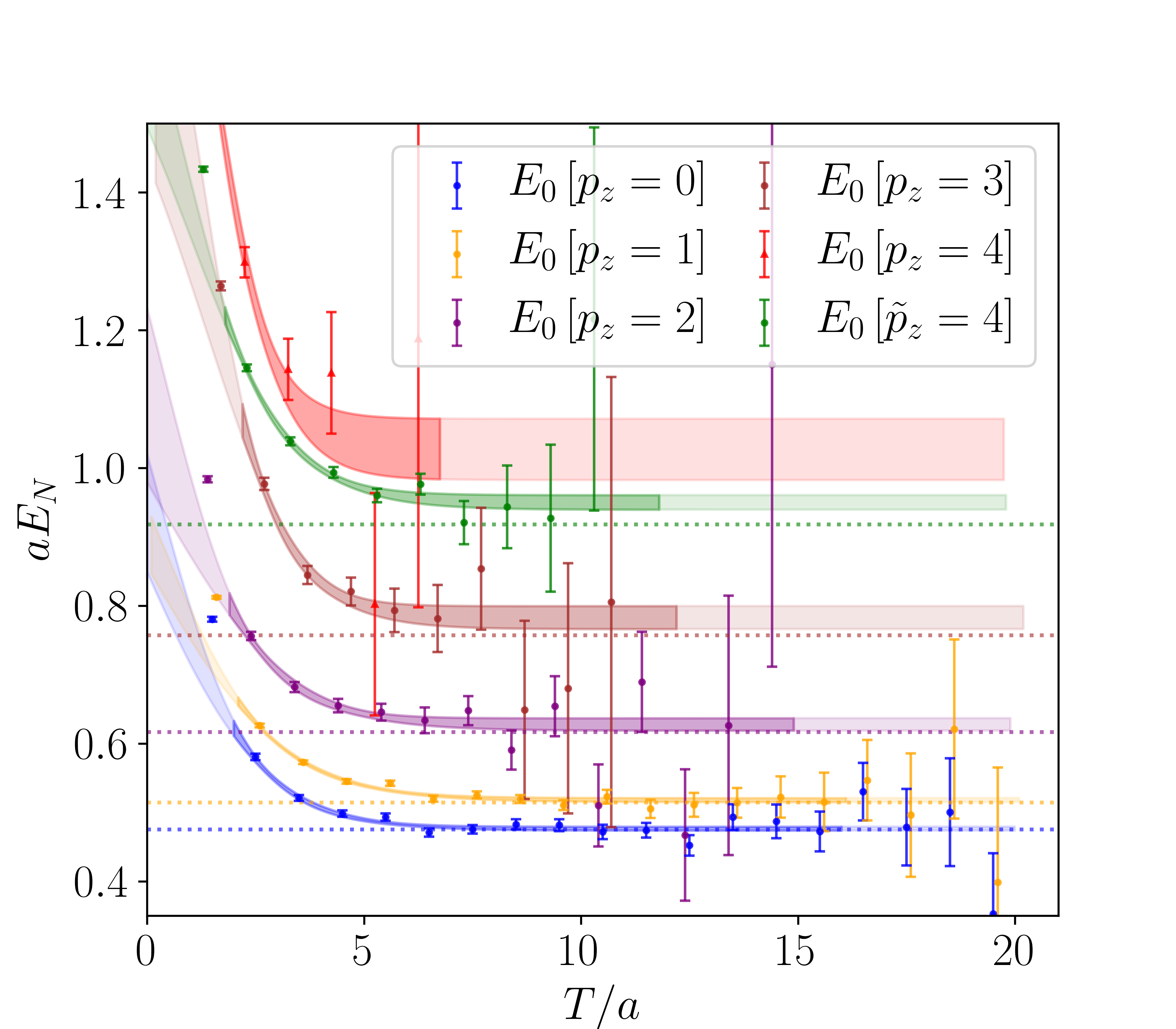}
  \caption{Nucleon effective energies for the $\Mlight$ ensemble using
    a projected interpolator obtained from the
    $\mathcal{B}_{\vec{p}=\vec{0}}$ and
    $\mathcal{B}_{\vec{p}\neq\vec{0}}$ bases subduced into the
    relevant little group, together with continuum expectations
    (dashed), and 2-state fits (bands), where in each case the darker
    region denotes the time series included in the fit.  No phasing
    was used to extract the ground-state nucleon energy for lattice
    momenta $ap_z\in\mathbb{Z}_5$, while $ap_z=4\left(2\pi/L\right)$
    was also determined with two units of phasing~\eqref{eq:twoPhase}.
    In the case of $ap_z=4\left(2\pi/L\right)$, the results
    with and without phased eigenvectors are shown as the green and
    red points respectively, clearly demonstrating the need for
    phasing. Data shifted for
    legibility, and shown for signal-to-noise ratios greater than
    1.35.\label{fig:effectiveenergies2390_proj}}
\end{figure}

We begin by presenting in Fig.~\ref{fig:effectiveenergies2390_proj}
the nucleon effective energies computed on the $\Mlight$ ensemble
using ground-state interpolating operators obtained from the
variational method with the $\mathcal{B}_{\vec{p}=\vec{0}}$ and
$\mathcal{B}_{\vec{p}\neq\vec{0}}$ bases, following the procedure
described for the $\Mheavy$ ensemble.  At all values of the momenta
shown (i.e. $ap_z\le4\left(2\pi/L\right)$) we show the results without
phasing; for $a p_z=4\left(2\pi/L\right)$, we also show the results
using the phased eigenvectors, as described above.  The need for
phasing at this value of the momenta (green) and above is striking, where the
plateau in the effective energy is clear at far greater temporal
separations, and the resulting energy far more precisely determined.
We observe that at such a lighter pion mass, the variational
method without phasing is insufficient to extract the ground-state
nucleon energy for $ap_z\geq4\left(2\pi/L\right)$ (red), but arguably
$ap_z\geq3\left(2\pi/L\right)$ (brown). We do not
expound further on nucleon energies for this
ensemble, however this demonstration underscores the need for
variational improvement of a phased distillation space in order to
study physical observables at high-momenta.

\subsection{Charges}
We isolate forward isovector matrix elements by constructing nucleon three-point functions
\begin{align}
  C_{3\text{pt}}\left(T,\tau,\vec{p}\right)&=\sum_{\vec{x},\vec{y},\vec{z}}e^{i\vec{p}\cdot\left(\vec{y}-\vec{x}\right)}\mathcal{P}_{\beta\alpha}^{3\text{pt}}\times \nonumber \\
  &\langle\mathcal{N}_\alpha\left(\vec{y},T\right)\mathcal{O}_\Gamma^{u-d}\left(\vec{z},\tau\right)\overline{\mathcal{N}}_\beta\left(x,0\right)\rangle,
  \label{eq:3ptfn}
\end{align}
with $\mathcal{O}_\Gamma^{u-d}$ an isovector insertion introduced at time $\tau$ between nucleon interpolators with temporal separation $T$, and $\mathcal{P}_{\beta\alpha}^{3\text{pt}}=\mathcal{P}^{2\text{pt}}\left(1+i\gamma_5\gamma_3\right)$ a $z$-polarized positive-parity projector. To study the asymptotic $0\ll\tau\ll T$ behavior, we parameterize our two-point and three-point~\eqref{eq:3ptfn} correlation functions according to 2-state fitting functionals
\begin{align}
  &C_{\text{fit}}^{2\text{pt}}\left(T\right)=e^{-E_0T}\left({\it{a}}+{\it{b}}e^{-\Delta ET}\right)\label{eq:2ptfit} \\
  &C_{\text{fit}}^{3\text{pt}}\left(T,\tau\right)=e^{-E_0T}\left(\mathcal{A}+\mathcal{B}e^{-\Delta ET}\right.\nonumber \\
  &\quad\quad\quad\quad\quad\left.+\mathcal{C}e^{-\Delta E\frac{T}{2}}\cosh\left[\Delta E\left(\tau-\frac{T}{2}\right)\right]\right)\label{eq:3ptfit},
\end{align}
where $\Delta E$ is the energy gap between the ground-state ($E_0$) and an effective first-excited ($E_1$) state, $\mathcal{B}$ and $\mathcal{C}$ respectively contain excited and transition matrix elements, and $\mathcal{A}$ contains the desired forward matrix element. Priors are again introduced to enforce the positivity of $\lbrace a,b\rbrace$. With these parametrizations, the desired ground-state matrix element is then $g_{00}^\Gamma=\mathcal{A}/a$ in the large-$T$ limit, as shown in~\cite{Egerer:2018xgu}. We perform simultaneous correlated fits to the computed two-point and three-point correlators according to~\eqref{eq:2ptfit} and~\eqref{eq:3ptfit} to extract these parameters. Contact terms arising from the fermion action are excluded from the simultaneous fits by fitting in the windows $\tau_{\text{fit}}/a\in\left[2,T-2\right]$ and $T_{\text{fit}}/a\in\left[2,T_{\text{fit}}^{\text{max}}\right]$, where $T_{\text{fit}}^{\text{max}}$ is set by the maximal temporal range for which the associated principal correlators have signal-to-noise ratios exceeding unity:
\begin{itemize}
\item $\lattMom=0$: $T_{\text{fit}}^{\text{max}}=16$
\item $\lattMom=1$: $T_{\text{fit}}^{\text{max}}=16$
\item $\lattMom=4$ - no phase: $T_{\text{fit}}^{\text{max}}=7$
\item $\lattMom=4$ - phased: $T_{\text{fit}}^{\text{max}}=12$.
\end{itemize}

When computing hadronic charges, the degree of excited-state contamination present in the three-point correlators for a given interpolator separation $T$ is often quantified (c.f.~\cite{Egerer:2018xgu,Yoon:2016jzj})  via definition of an effective charge
\begin{equation*}
  g_{\text{eff}}^\Gamma\left(T,\tau\right)=C_\Gamma^{3\text{pt}}\left(T,\tau\right)/C_{\text{fit}}^{2\text{pt}}\left(T\right),
\end{equation*}
where the numerator is a three-point correlation function with inserted Dirac structure $\Gamma$ computed for intermediate times $\tau/a=\left[0,T-1\right]$, and $C_{\text{fit}}^{2\text{pt}}\left(T\right)$ is the two-point function fit evaluated at the source-sink interpolator separation $T$. This ratio has the advantage of plateauing to $g^\Gamma_{00}$ as $\tau$ and $T-\tau$ become large, but is only useful in so far as $C_{\text{fit}}^{2\text{pt}}$ is well-determined and sufficiently captures the ground-state. We find this ratio, particularly in the high-momentum frames considered, to be misleading when juxtaposed with the ratio of the simultaneous $C_\Gamma^{3\text{pt}}\left(T,\tau\right)$ and $C_{\text{fit}}^{2\text{pt}}\left(T\right)$ fit. We instead illustrate the quality of our data by forming a direct ratio of the computed correlation functions
\be
R_\Gamma\left(T,\tau\right)=C_\Gamma^{3\text{pt}}\left(T,\tau\right)/C^{2\text{pt}}\left(T\right).
\label{eq:ratio3pt2pt}
\ee
All following figures depict these ratios~\eqref{eq:ratio3pt2pt} together with ratios of the fitted three-point and two-point functions for each $T/a$, as well as the extracted renormalized isovector charge indicated with a black line and grey errorband. Data excluded from fits are in grey. All errors are determined via a simultaneous jackknife resampling of the data.

\section{Charge Behavior\label{sec:chargeBehavior}}
\subsection{$g_S^{u-d}$}
\begin{figure*}[tb]
  \centering
  \subfigure[]{\includegraphics[width=0.49\linewidth]{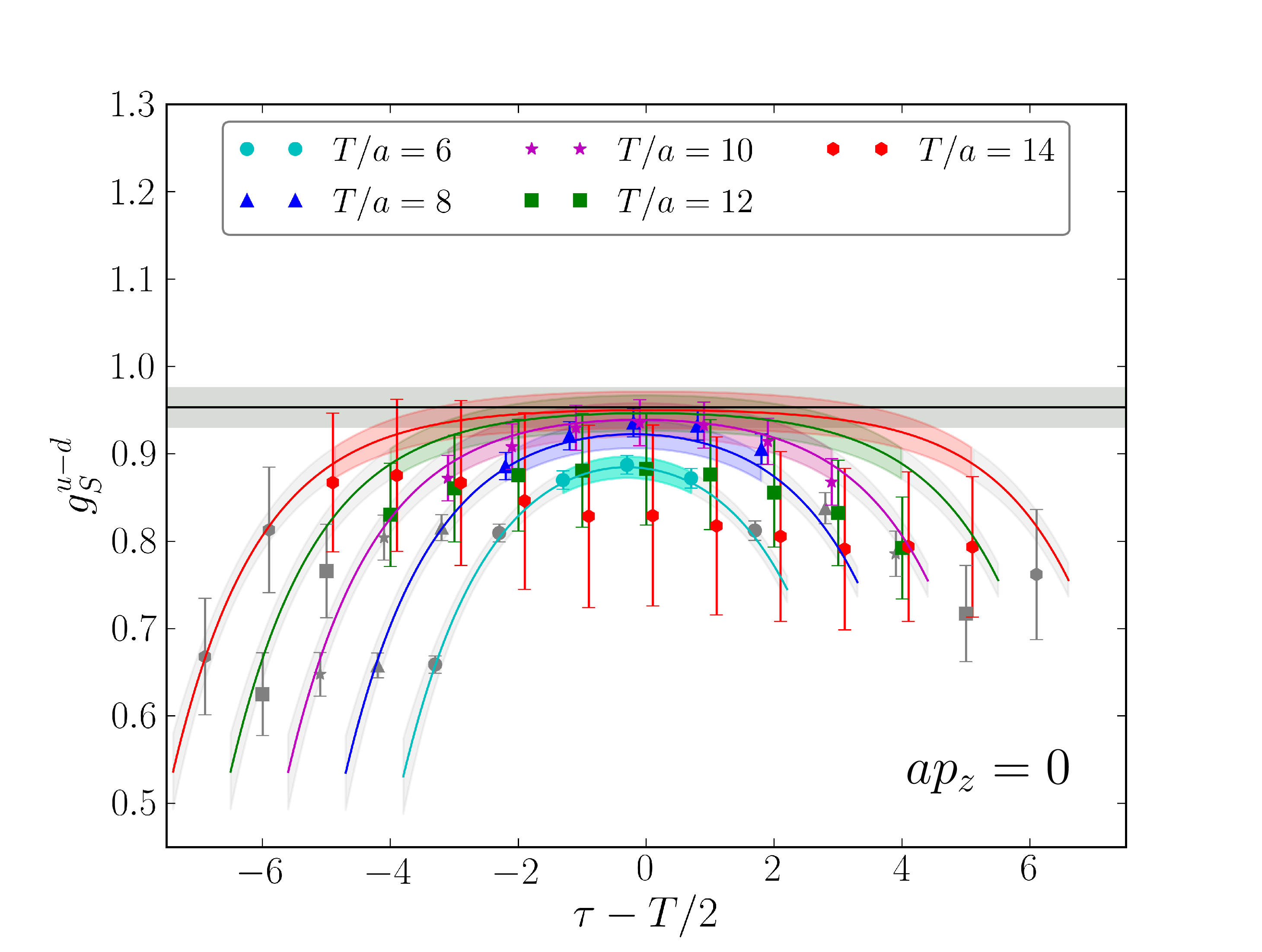}\label{fig:scalar_p000}}
  \subfigure[]{\includegraphics[width=0.49\linewidth]{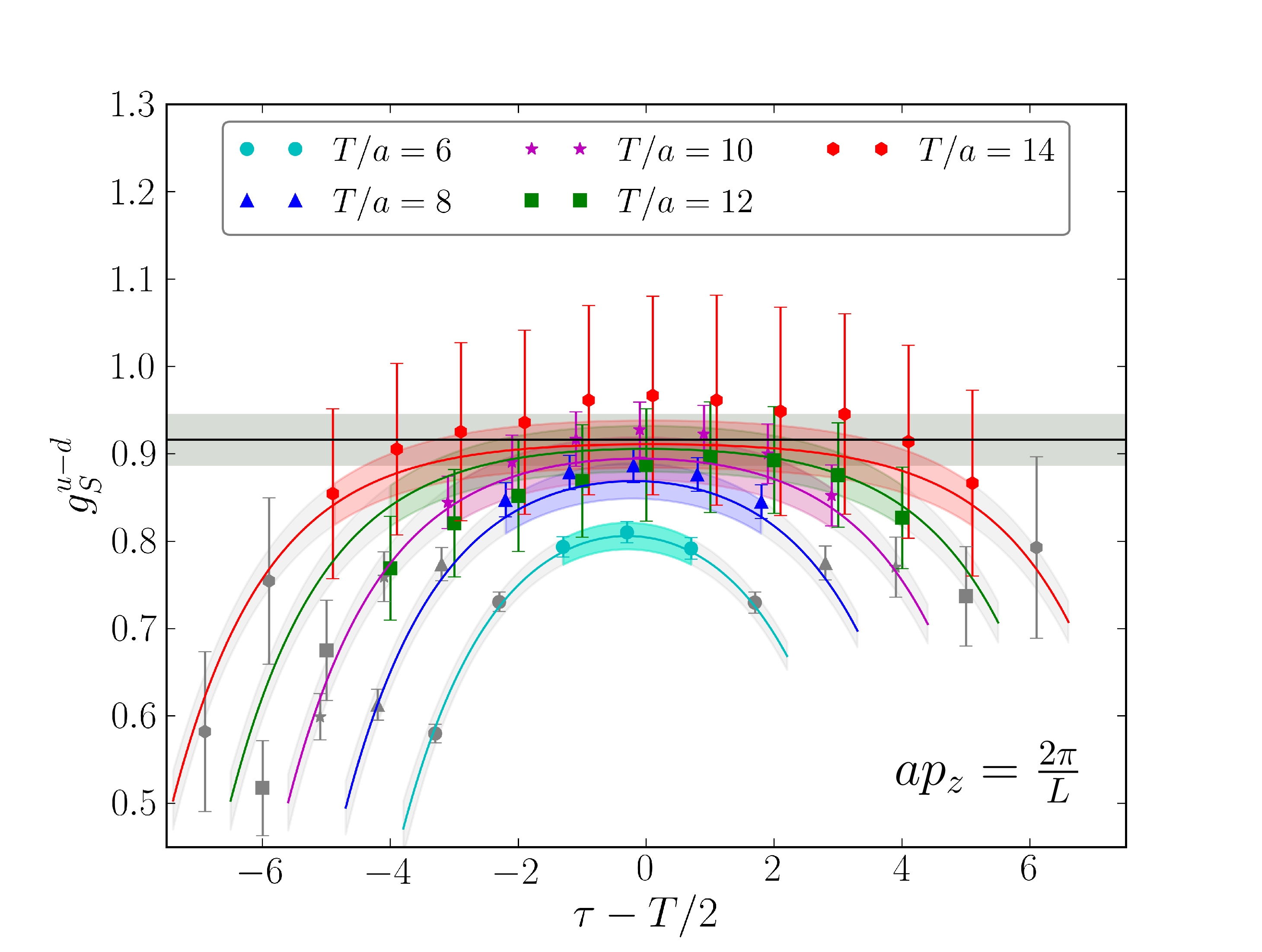}\label{fig:scalar_p001}}
  \subfigure[]{\includegraphics[width=0.49\linewidth]{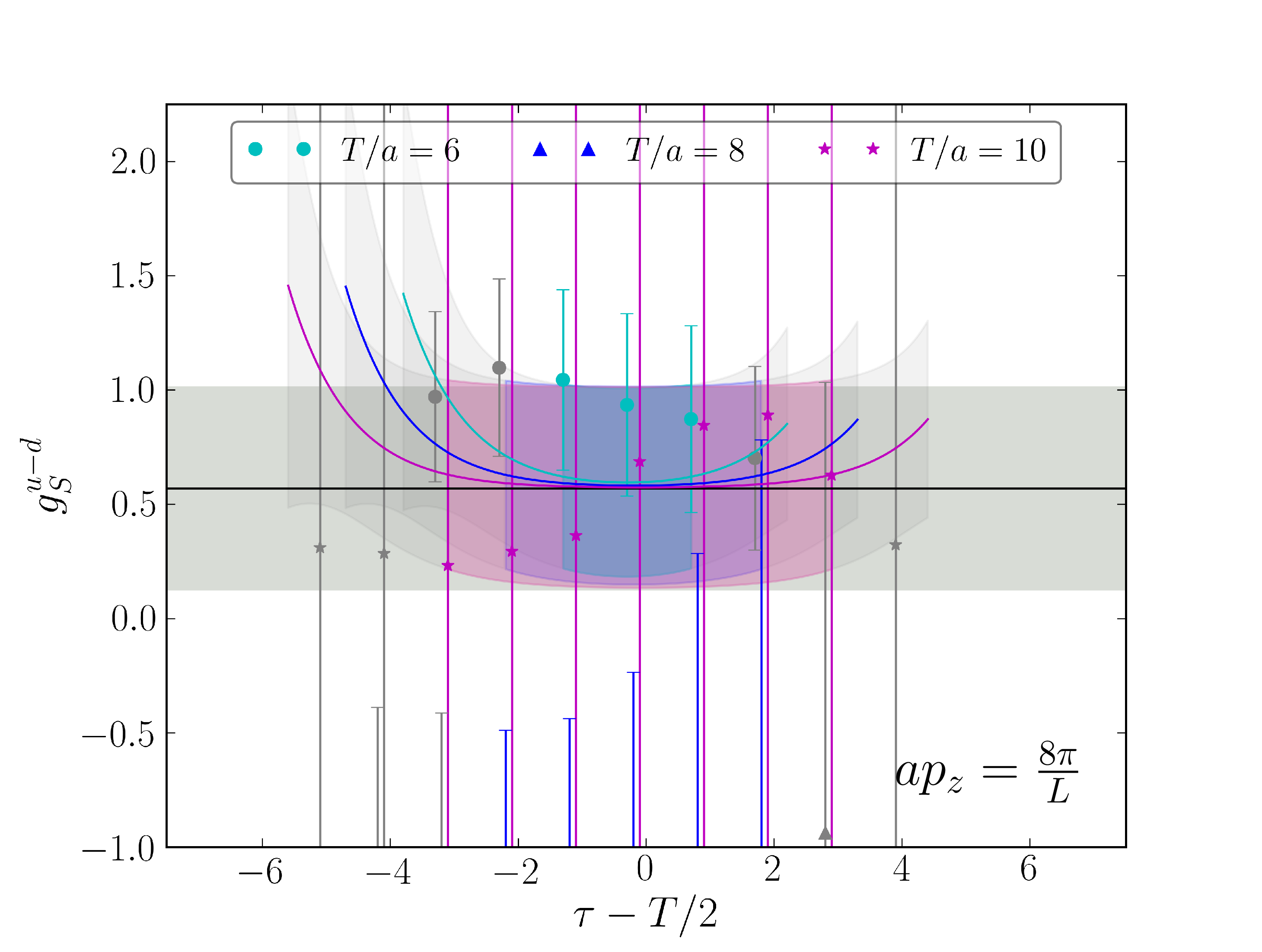}\label{fig:scalar_p004_nophase}}
  \subfigure[]{\includegraphics[width=0.49\linewidth]{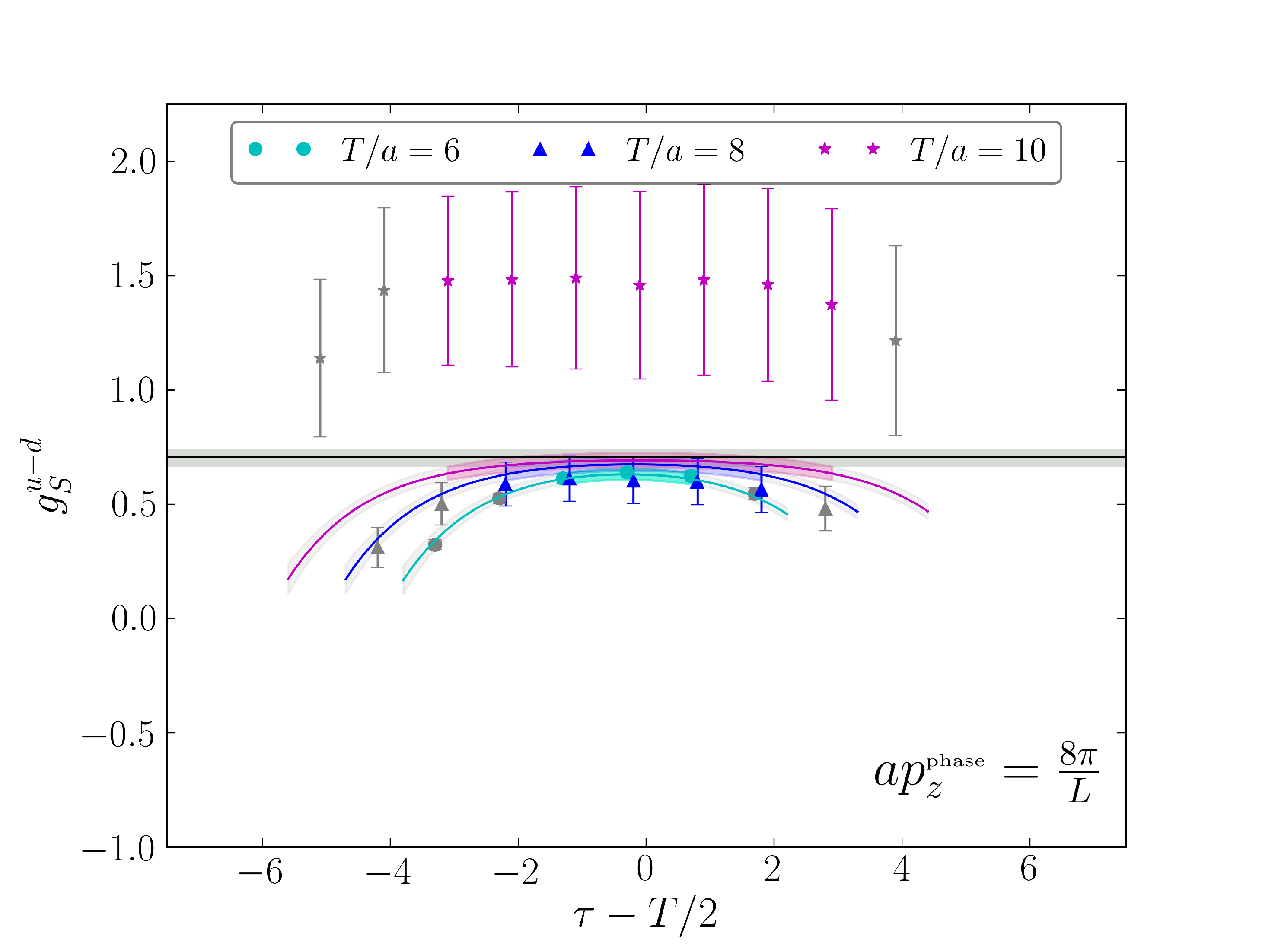}\label{fig:scalar_p004_phase}}
  \caption{Extracted renormalized $R_S\left(T,\tau\right)$ and isovector scalar charges for momenta (a) $ap_z=0$, (b) $ap_z=\left(2\pi/L\right)$, (c) $ap_z=4\times\left(2\pi/L\right)$ without phasing, and (d) $ap_z=4\times\left(2\pi/L\right)$ with two units of allowed lattice momentum applied to eigenvectors. Variationally improved operators were used within each momentum channel.\label{fig:scalar}}
\end{figure*}

The isovector scalar $S=\overline{q}\frac{\tau^3}{2}q$ current within nucleon states decomposes trivially as
\be
\bra{N}S\ket{N}=\frac{1}{2M_N}\overline{u}_N\left(p_f\right)G_S^{u-d}\left(q^2\right)u_N\left(p_i\right),
\ee
where $G_S^{u-d}$ is the isovector scalar form factor. The amplitude $G_S^{u-d}$ is Lorentz-invariant and should thus be independent of the nucleon boost, absent excited-state, discretization and finite-volume effects. In particular, in the forward limit one should, in principle, be able to access $G_S^{u-d}\left(0\right)=g_S^{u-d}$ regardless of frame. Figure~\ref{fig:scalar} illustrates the $R_S\left(T,\tau\right)$ ratios needed to access the scalar charge and associated fits within our considered nucleon frames, demonstrating the degree to which this supposition is realized. In the rest frame a clear plateau is observed in the ratio by $T/a=10$, while determinations at larger values of $T/a$ deviate from this trend and exhibit increased uncertainty; the latter being consistent with the observed variability of the nucleon effective energies at these same times. Most notable is a reduction in value and uncertainty of $g_S^{u-d}$ when compared with standard, high statistics, smearing schemes on the same $\Mlight$ ensemble. Namely in~\cite{Yoon:2016jzj}, it was found $g_S^{u-d}=0.990(89)$ - the use of distillation has led to a more precise determination by $\sim75\%$.

Considering the $ap_z=\left(2\pi/L\right)$ frame, we observe statistical consistency with the $ap_z=0$ determination, with a plateau emerging for $T/a\sim10-12$. The expected increase of excited-state contamination is evident in Fig.~\ref{fig:scalar_p001}, where there exists greater curvature of the ratio data for a given $T$ and the difference between each $R_S\left(T,\tau\right)$ plateau and the asymptotic charge is seen to increase relative to the rest case. This amounts to marked increases in $\mathcal{B}$ and $\mathcal{C}$ of Eq.~\ref{eq:3ptfit} which capture excited-state $\bra{N'}S\ket{N'}$ and transition $\bra{N'}S\ket{N}$ matrix elements, respectively.

Without introduction of appropriate momentum phases into the distillation space, attempts to access the scalar charge in a highly-boosted frame are utterly meaningless (Fig.~\ref{fig:scalar_p004_nophase}).
Isolation of the scalar charge in the $ap_z=4\times\left(2\pi/L\right)$ frame is however dramatically improved when a phased distillation space is used. The statistical precision of the $R_S\left(T,\tau\right)$ data improves considerably, provided the two-point function is well-determined. However, the extracted charge is dubious - the phased determination differs by $25\%$ from the average of the $ap_z=\{0,1\}\times\left(2\pi/L\right)$ cases. The close proximity of the $R_S\left(T,\tau\right)$ plateaus for each $T/a$ and the asymptotic charge suggest that at the level of the 2-state fits considered herein, the first excited-state matrix element is small. However, without performing $R_S\left(T,\tau\right)$ computations for additional $T/a$ and performing higher state fits, this cannot be rigorously confirmed. We do point out the statistical noise evident in the $T/a=10$ data is not surprising, as the phased two-point function loses signal at $T/a\sim10$ (c.f. Fig.~\ref{fig:effectiveenergies2390_proj}). Furthermore, determinations of $Z_S$ found in~\cite{Yoon:2016jzj} vary below the $2\%$ level and thus also cannot explain the observed discrepancy. One may be tempted to attribute this dramatic difference to a mixing of the scalar current with the derivative of the vector current $D_\mu\lbrace\overline{\psi}\gamma_\mu\psi\left(x\right)e^{-iq\cdot x}\rbrace$. Given the explicit zero 3-momentum transfer with the probing current, it is evident this derivative mixing is only possible for $q_4\neq0$ or when unwanted excited-to-ground state transitions are present. This possibility is captured by $\mathcal{C}$ of~\eqref{eq:3ptfit}, and is reflected in the overall curvature of $R_S\left(T,\tau\right)$ rather than vertical shifts of the computed matrix element. We are left to attribute this puzzling discrepancy to statistical fluctuations and the lack of additional $T/a$ data. As will be shown, the other charges we explore exhibit much greater consistency in the studied momentum frames, and observed deviations can be attributed to known systematic effects. Table.~\ref{tab:scalarTable} catalogs the isolated scalar charges and the correlated figure of merit for the simultaneous fits of each frame.

\begin{table}[t]
  \begin{center}
    \setlength\abovecaptionskip{-1pt}
    \setlength{\belowcaptionskip}{-10pt}
    \begin{tabular}{c|c|c|c|c}\hline\hline
      $g_\Gamma$ & $ap_z=0$ & $ap_z=2\pi/L$ & $ap_z=8\pi/L$ & $ap_z^{\text{phase}}=8\pi/L$ \\
      \hline
      $g_S^{u-d}$ & 0.953(22) & 0.916(28) & 0.57(44) & 0.705(35) \\
      $\chi^2_r$ & 0.920 & 1.010 & 12.482 & 2.037 \\
      \hline\hline
    \end{tabular}
  \end{center}
  \caption{Renormalized isovector scalar charge determined at rest and in boosted frames.\label{tab:scalarTable}}
\end{table}

\subsection{$g_V^{u-d}$}
\begin{figure}[h!]
  \centering
  \includegraphics[width=\linewidth]{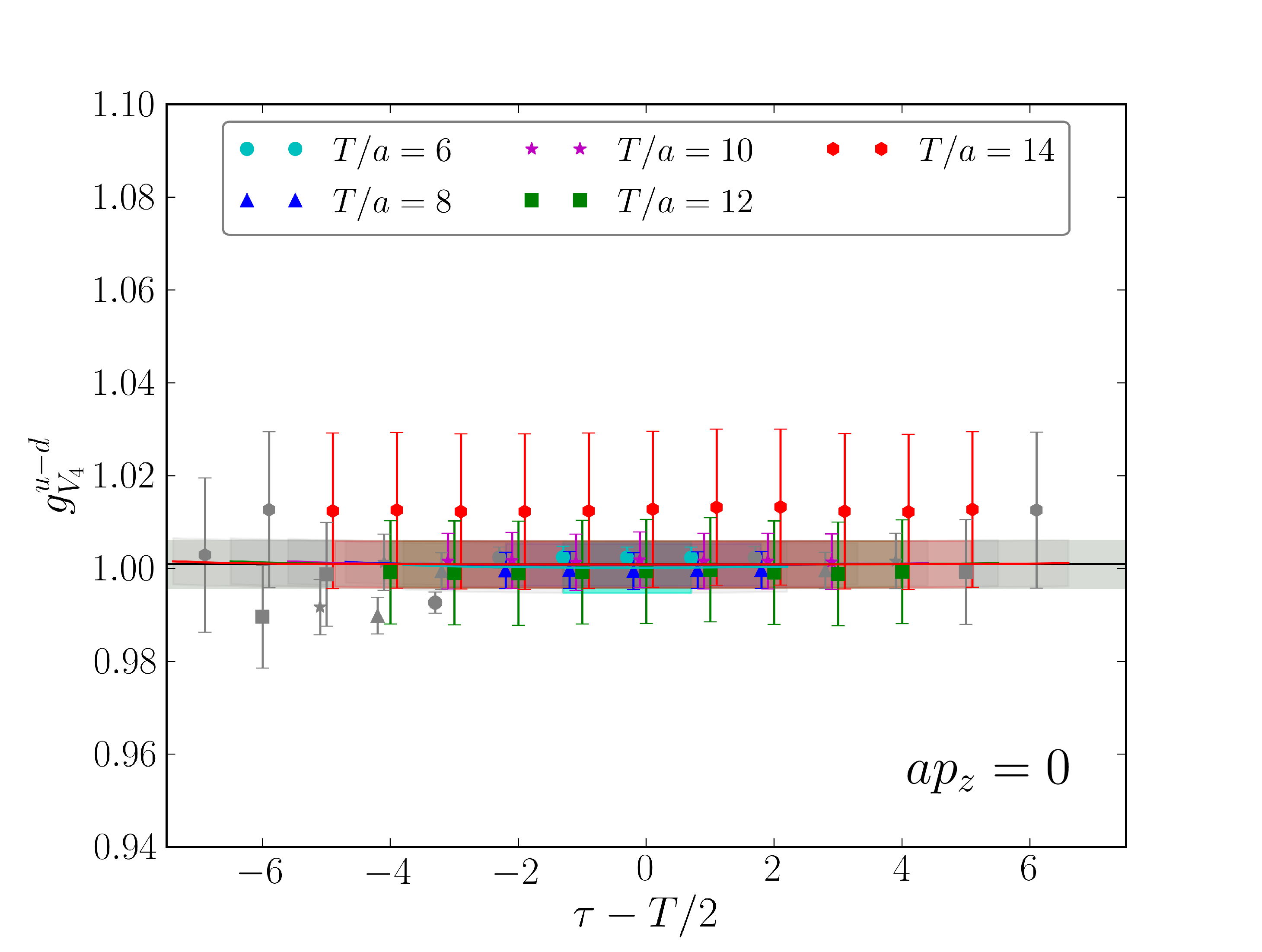}
  \caption{Extracted renormalized $R_{V_4}\left(T,\tau\right)$ and isovector vector charges determined for momenta $ap_z=0$. A variationally improved operator was used in these determinations.\label{fig:vector4_p000}}
\end{figure}

\begin{figure*}[tb]
  \centering
  \subfigure[]{\includegraphics[width=0.49\linewidth]{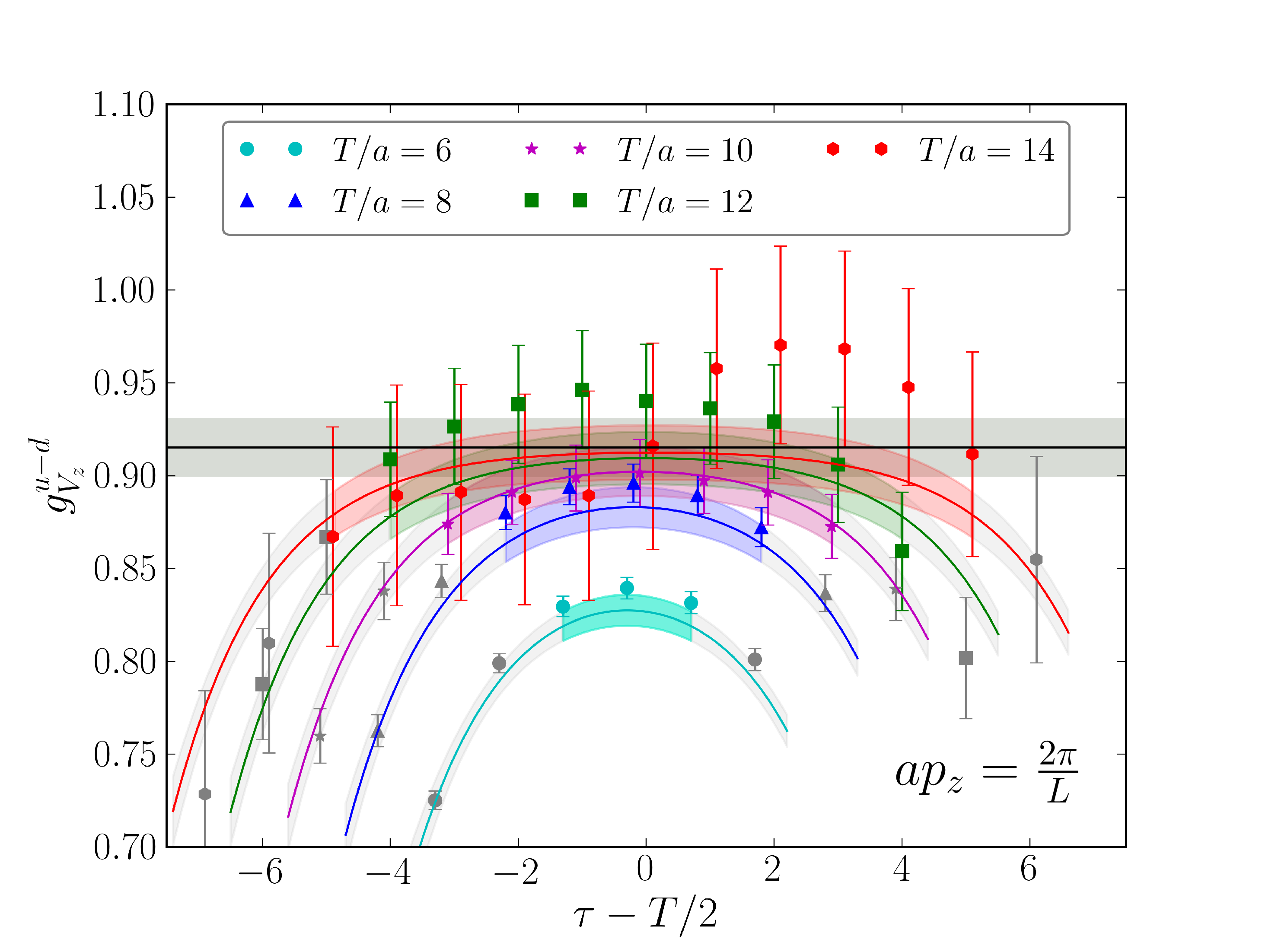}\label{fig:vectorZ_p001}}
  \subfigure[]{\includegraphics[width=0.49\linewidth]{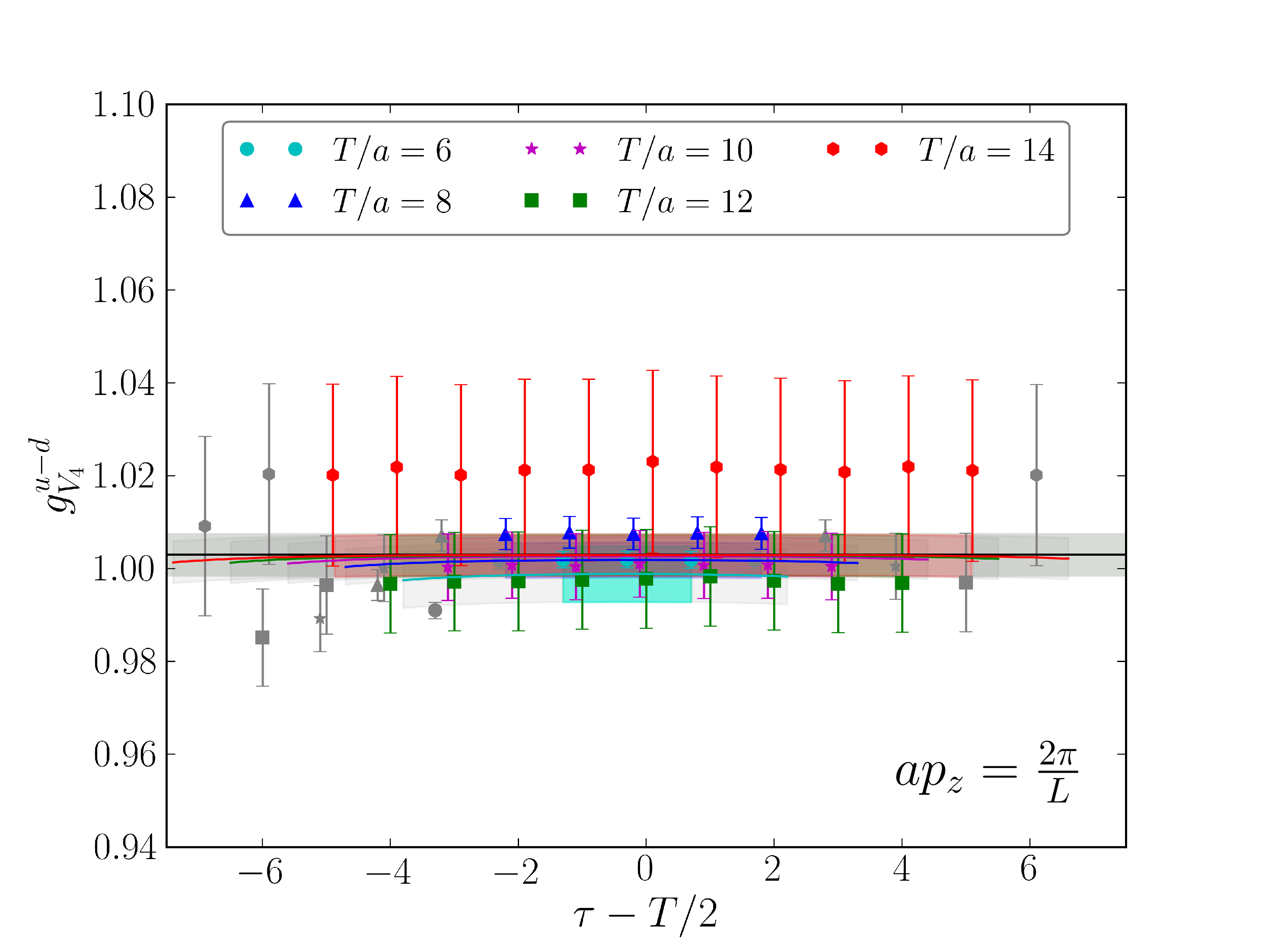}\label{fig:vector4_p001}}
  \subfigure[]{\includegraphics[width=0.49\linewidth]{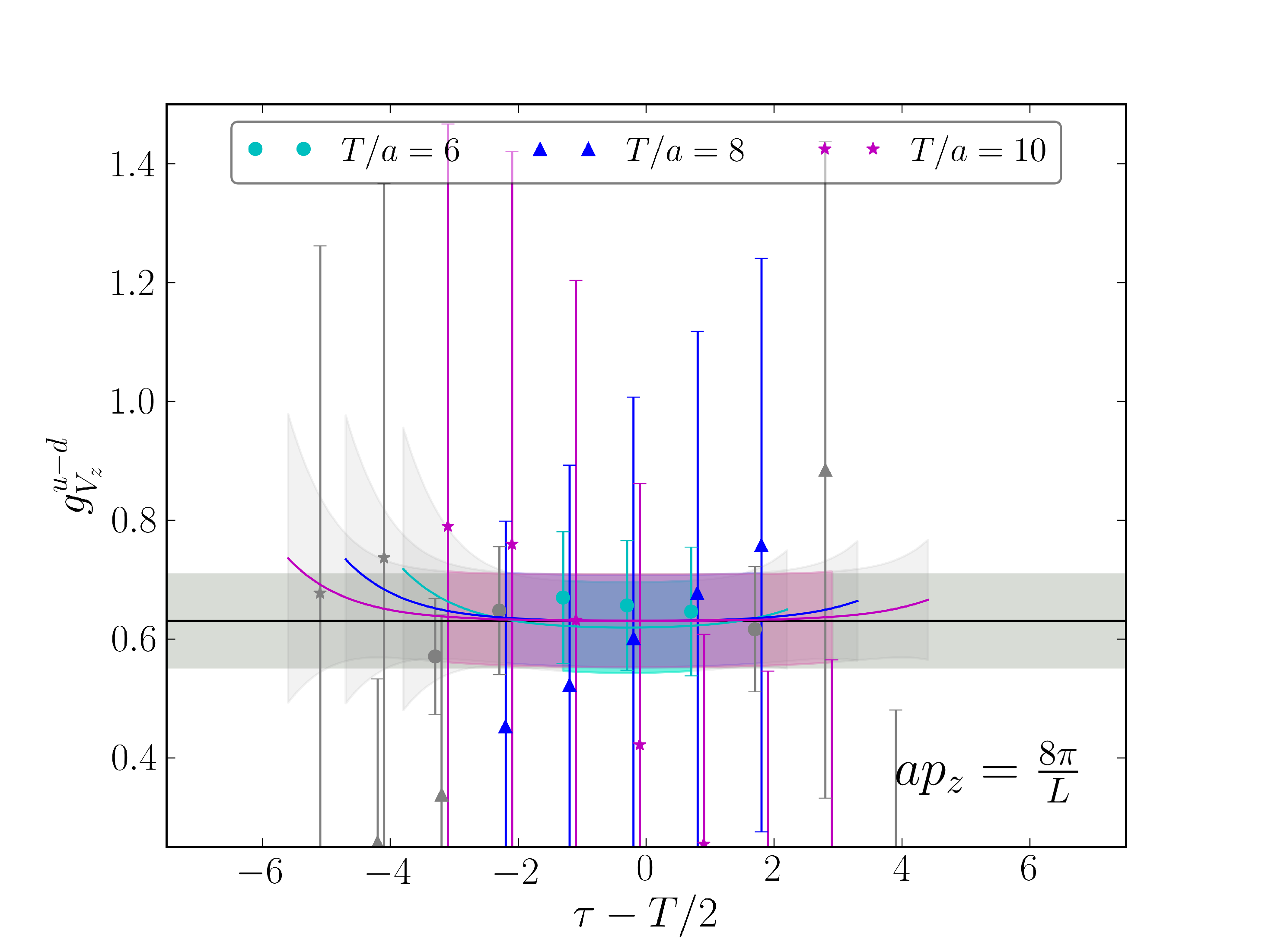}\label{fig:vectorZ_p004_nophase}}
  \subfigure[]{\includegraphics[width=0.49\linewidth]{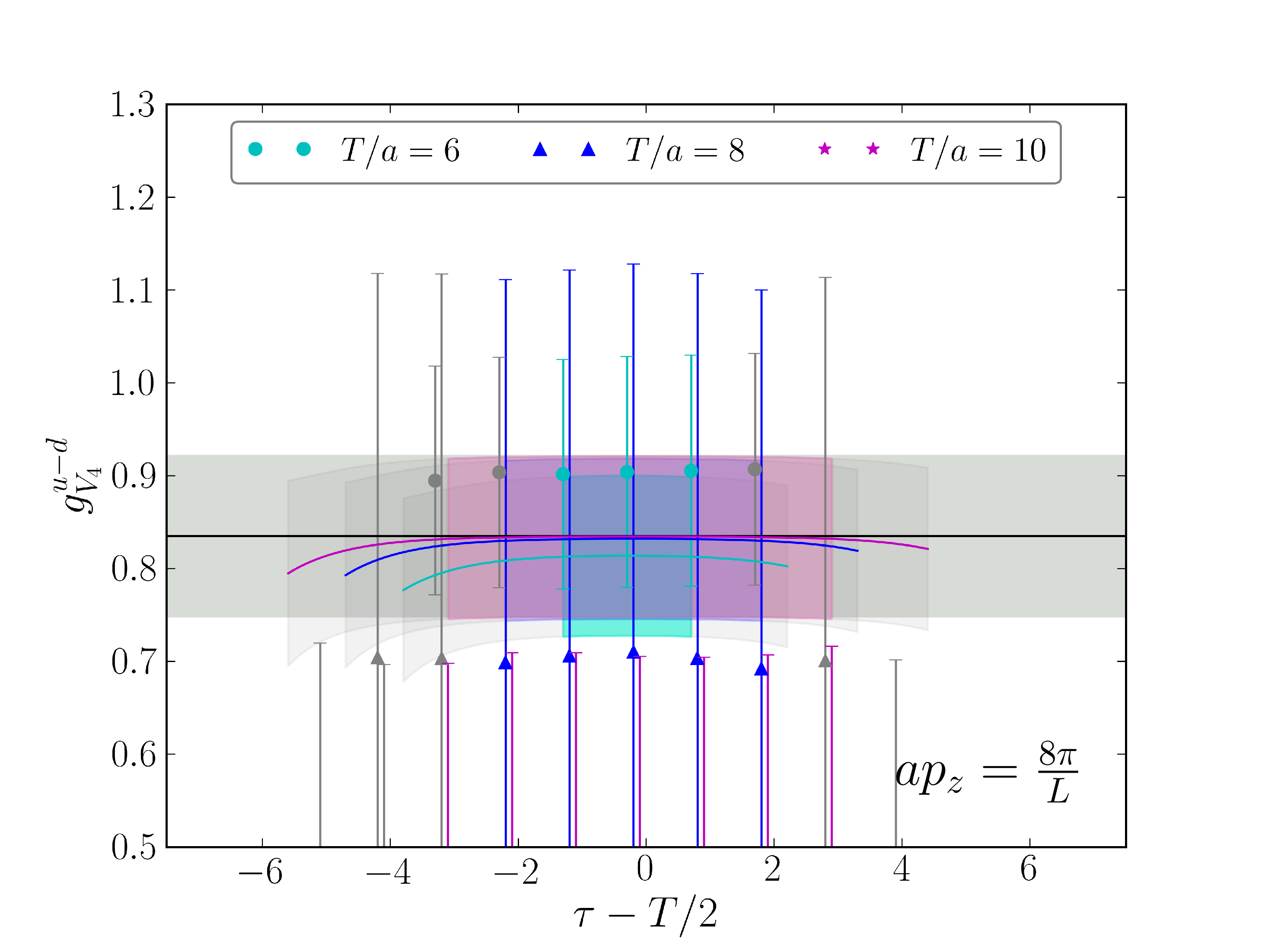}\label{fig:vector4_p004_nophase}}
  \subfigure[]{\includegraphics[width=0.49\linewidth]{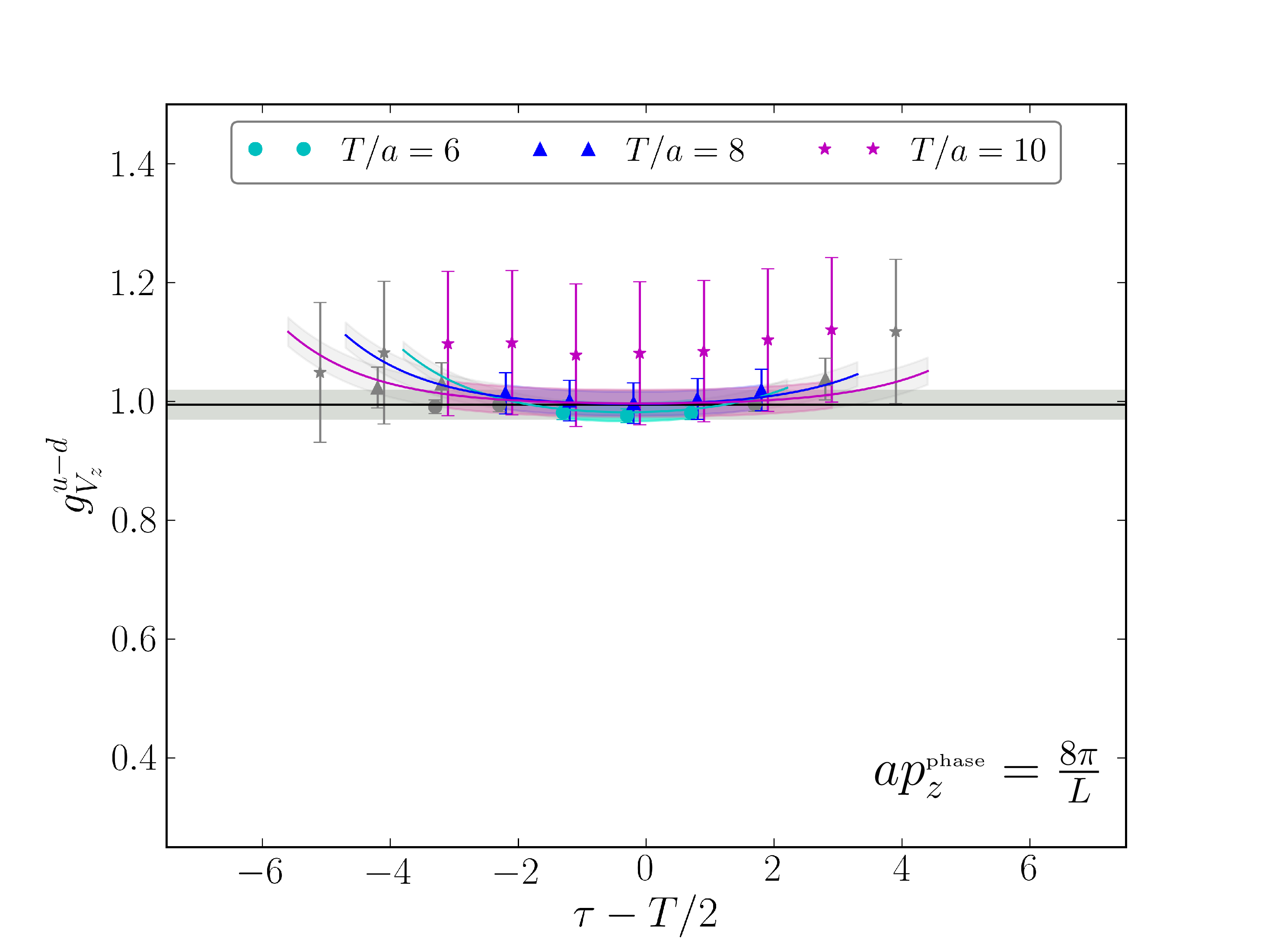}\label{fig:vectorZ_p004_phase}}
  \subfigure[]{\includegraphics[width=0.49\linewidth]{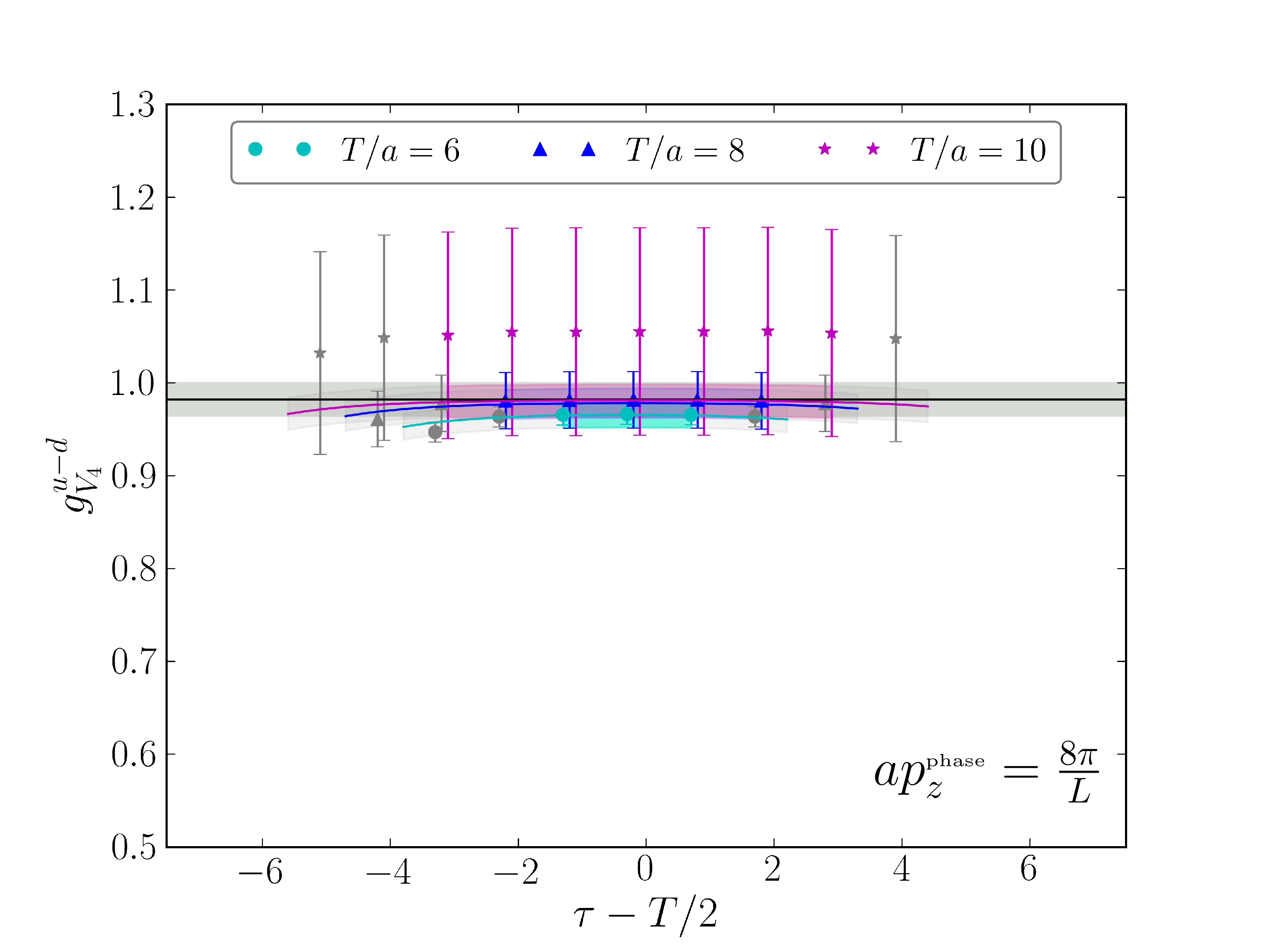}\label{fig:vector4_p004_phase}}
  \caption{Extracted renormalized $R_{V_\mu}\left(T,\tau\right)$ and isovector vector charges determined from $\gamma_3$ (left panel) and $\gamma_4$ (right panel) insertions. External nucleon momentum according to {(a),(b)} $ap_z=\left(2\pi/L\right)$, {(c),(d)} $ap_z=4\left(2\pi/L\right)$ without phasing, {(e),(f)} $ap_z=4\left(2\pi/L\right)$ with two units of allowed lattice momenta applied to eigenvectors. Variationally improved operators were used within each momentum channel.}
\end{figure*}

Among the currents considered, the vector current $V_\mu=\overline{q}\gamma_\mu\frac{\tau^3}{2}q$ is unique given that it is a conserved quantity in the continuum. Our decision to adopt purely local currents in this work necessarily violates this conservation. However the derived vector current renormalization constant~\cite{Yoon:2016jzj} reestablishes the desired conservation up to quadratic corrections in the lattice spacing - namely, $Z_Vg_{V,{\text{bare}}}^{u-d}=1+\mathcal{O}\left(a^2\right)$. Considering the vector current Lorentz structure between the ground-state nucleon and an arbitrary state $N'$ with nucleon quantum numbers
\begin{align*}
  \bra{N'}V_\mu\ket{N}&=\overline{u}_{N'}\left(p_f\right)\left[F_1^{u-d}\left(q^2\right)\left(\gamma_\mu-\frac{q_\mu}{q^2}\slashed{q}\right)+\right. \\
    &\left.\frac{\sigma_{\mu\nu}q_\nu}{M_{N'}+M_N}F_2^{u-d}\left(q^2\right)\right]u_{N}\left(p_i\right),
\end{align*}
it is clear for $\vec{q}=0$ the temporal component of the vector current simply yields the baryon number of the nucleon and all its excitations. A useful sanity check then for the phasing considered herein, is to ensure the renormalized $g_V^{u-d}$ is unity in the $V_4=\overline{q}\gamma_4q$ channel for each forward frame considered. As illustrated in Figures~\ref{fig:vector4_p000},\ref{fig:vector4_p001} \& ~\ref{fig:vector4_p004_phase}, we indeed find $Z_V g^{u-d}_{V_4,\text{bare}}$ to be unity and
temporally invariant, most notably even as the nucleon momentum is increased and phasing is employed. A highly-boosted nucleon interpolator without phasing exhibits poor overlap with the ground-state nucleon (Fig.~\ref{fig:vector4_p004_nophase}) and is sufficiently noisy such that $Z_Vg_{V_4,\text{bare}}^{u-d}\neq1$. The extracted $g_{V_4}^{u-d}$ are presented in Tab~\ref{tab:vector4Table}, with consistent determinations observed in the $ap_z=\{0,1\}\times\left(2\pi/L\right)$ and $ap_z^{\text{phase}}=4\left(2\pi/L\right)$ momentum channels.
\begin{table}[b]
  \begin{center}
    \setlength\abovecaptionskip{-1pt}
    \setlength{\belowcaptionskip}{-10pt}
    \begin{tabular}{c|c|c|c|c}\hline\hline
      $g_\Gamma$ & $ap_z=0$ & $ap_z=2\pi/L$ & $ap_z=8\pi/L$ & $ap_z^{\text{phase}}=8\pi/L$ \\
      \hline
      $g_{V_4}^{u-d}$ & 1.001(5) & 1.003(4) & 0.84(9) & 0.982(18) \\
      $\chi^2_r$ & 0.901 & 1.767 & 12.317 & 1.902 \\
      \hline\hline
    \end{tabular}
  \end{center}
  \caption{Renormalized isovector vector charges determined via $\gamma_4$ at rest and in boosted frames.\label{tab:vector4Table}}
\end{table}

Non-zero nucleon momenta while still with $\vec{q}=0$, opens the $V_z=\overline{q}\gamma_3\frac{\tau^3}{2}q$ channel as an additional means to quantify the ground-state Dirac form factor $F_1^{u-d}\left(0\right)$. However, any such attempt to isolate the ground-state Dirac form factor $F_1^{u-d}$ signal will be contaminated with the transition form factor $F_2^{u-d}\left(q^2\right)$ signal in proportion to $q_4/\left(M_{N'}+M_N\right)$. In the ideal scenario that excited states are completely removed, the energy transfer $q_4$ will vanish and $F_1^{u-d}\left(0\right)$ can be directly accessed with $V_z$. Figure~\ref{fig:vectorZ_p001} illustrates $R_{V_z}\left(T,\tau\right)$, which features a clear dependence on $\lbrace T,\tau\rbrace$ and whose asymptotic limit differs from $g_{V_4}^{u-d}$ by $\sim8\%$, together indicating the presence of excited-states. Thus absent a dedicated study and subsequent removal of the $F_2^{u-d}$ contamination, the best we can extract here is $F_1^{u-d}\left(q^2\right)-\tfrac{q_4\gamma_4}{M_{N'}+M_N}F_2^{u-d}\left(q^2\right)$ - which we will denote as $g_{V_z}^{u-d}$ for brevity. To the extent this pollution is unchanging in other forward frames is bore out in Figs.~\ref{fig:vectorZ_p004_nophase} \&~\ref{fig:vectorZ_p004_phase}. As for the scalar charge, the unphased $ap_z=4\left(2\pi/L\right)$ determination is meaningless and is dominated by uncertainty in the unaltered two-point function. The phased $ap_z=4\left(2\pi/L\right)$ determination, although statistically consistent with $g_{V_4}^{u-d}$, is constrained by only two values of $T$ and is characterized by a curious flip in concavity of $R_{V_z}\left(T,\tau\right)$. As this dependence is captured by $\mathcal{C}$ of Eq.~\ref{eq:3ptfit}, it is clear the effect of phasing has apparently identified the conjugate of the ground-to-first-excited state transition. This behavior warrants repeated calculations for additional values of $T/a$ with increased statistics to elucidate whether this behavior is merely fluctuations or a clear trend. That said, the $R_{V_z}\left(T,\tau\right)$ appears to be trending below unity within the well-determined values of $T/a$. Results of these simultaneous fits are cataloged in Tab.~\ref{tab:vectorZTable}.
\begin{table}[t]
  \begin{center}
    \setlength\abovecaptionskip{-1pt}
    \setlength{\belowcaptionskip}{-10pt}
    \begin{tabular}{c|c|c|c|c}\hline\hline
      $g_\Gamma$ & $ap_z=0$ & $ap_z=2\pi/L$ & $ap_z=8\pi/L$ & $ap_z^{\text{phase}}=8\pi/L$ \\
      \hline
      $g_{V_z}^{u-d}$ & -- & 0.915(15) & 0.63(8) & 0.995(23) \\
      $\chi^2_r$ & -- & 1.216 & 12.544 & 2.150 \\
      \hline\hline
    \end{tabular}
  \end{center}
  \caption{Renormalized $g_{V_z}^{u-d}$ determined via $\gamma_3$ in boosted frames. By definition, $g_{V_z}^{u-d}=0$ at rest.\label{tab:vectorZTable}}
\end{table}

\subsection{$g_A^{u-d}$}
\begin{figure}[t]
  \includegraphics[width=\linewidth]{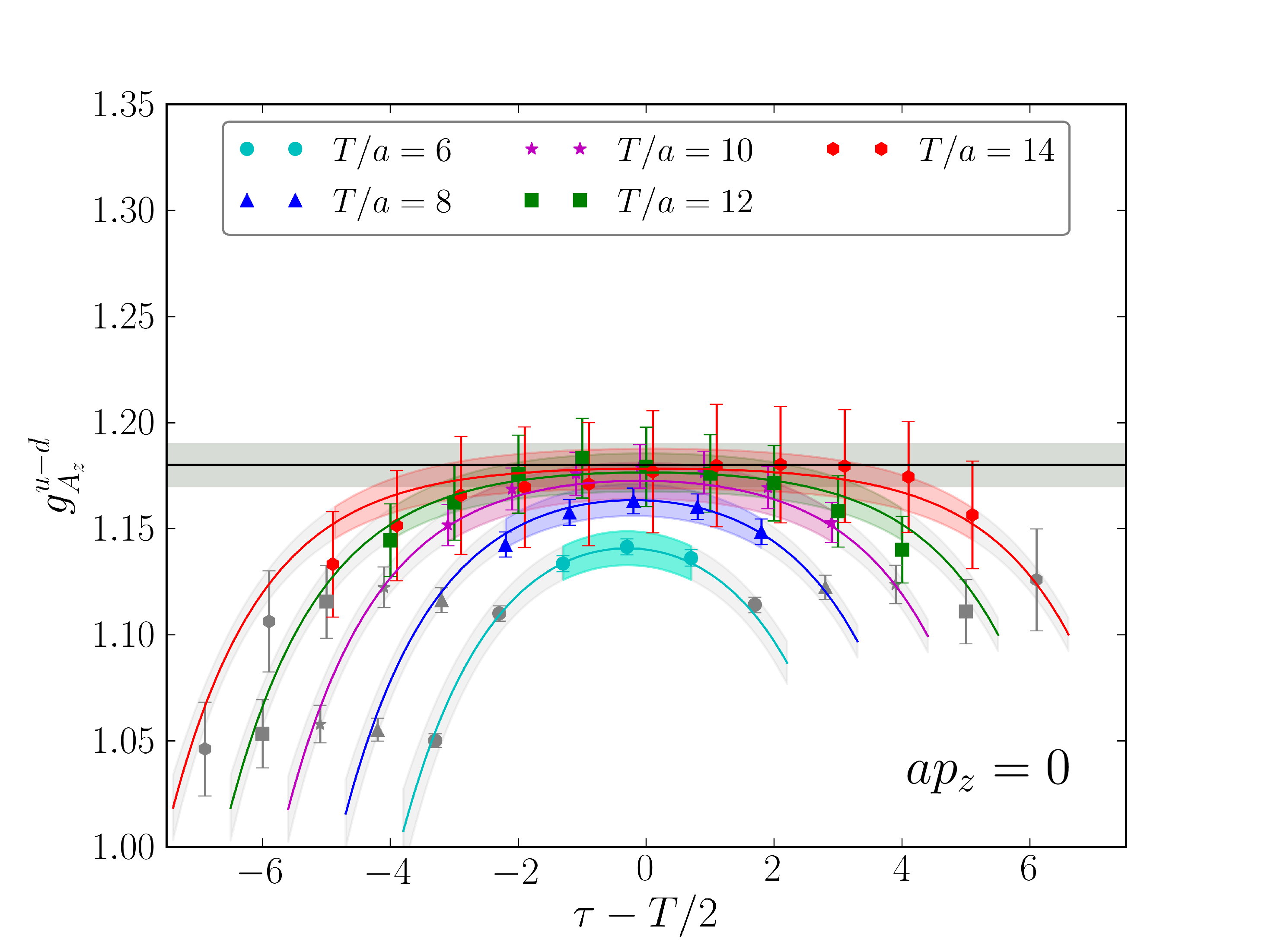}
  \caption{Extracted renormalized $R_{\gamma_3\gamma_5}\left(T,\tau\right)$ and isovector axial charge determined via $\gamma_3\gamma_5$ at rest. A variationally optimized operator was used in these determinations.\label{fig:axial11_p000}}
\end{figure}

The axial charge of the nucleon is perhaps the most enigmatic of the isovector charges given its long history as a benchmark in LQCD, and only recent efforts falling to within $1\%$ of experiment~\cite{Chang:2018uxx,Bali:2014nma,Horsley:2013ayv}. At zero-momentum the nucleon expectation of the axial current is vanishing except for components along the direction of polarization. Thus for our $z$-polarized nucleons, we must use $\gamma_3\gamma_5$ at rest to access $g_A^{u-d}$ - which we denote as $g_{A_z}^{u-d}$. Together with a Lorentz decomposition of the axial current
\begin{align}
  &\bra{N}A_\mu\ket{N}=\overline{u}_N\left(p_f\right)\left[\gamma_\mu\gamma_5G_A^{u-d}\left(q^2\right)-\right.\nonumber  \\
    &\left.i\frac{q_\mu}{2M_N}\gamma_5\widetilde{G}_P^{u-d}\left(q^2\right)\right]u_N\left(p_i\right)
  \label{eq:axialDecomp}
\end{align}
and $\vec{q}=0$, it is evident the axial matrix element at rest receives contributions only from the axial form factor and not the induced pseudoscalar form factor. We plot in Fig.~\ref{fig:axial11_p000} the renormalized $R_{\gamma_3\gamma_5}\left(T,\tau\right)$ and $g_{A_z}^{u-d}$ isolated at rest from our simultaneous fits. We observe noticeable contamination from excited-states for $T/a=\{6,8\}$, but broad consistency for the remaining $T/a$ values. The observed $\sim7\%$ deviation from the experimental value of $1.2756(13)$~\cite{PDG} is not the focus of this work, but is conventionally attributed to finite-volume effects and excited-states. In fact, it has been observed~\cite{Jang:2019vkm} that the $\gamma_3\gamma_5$ channel is particularly sensitive to closely-spaced excited states, which when incorrectly identified leads to not only a discrepancy of $g_{A_z}^{u-d}$ with experiment but also a violation of the operator derived PCAC relation. Our deviation of $g_{A_z}^{u-d}$ from experiment is, however, consistent with~\cite{Yoon:2016jzj}, where with standard smearing schemes on the same $\Mlight$ ensemble it was found $g_{A_z}^{u-d}=1.208(33)$. We emphasize the use of distillation has led to a 3-fold reduction in uncertainty.
\begin{table}[b]
  \begin{center}
    \setlength\abovecaptionskip{-1pt}
    \setlength{\belowcaptionskip}{-10pt}
    \begin{tabular}{c|c|c|c|c}\hline\hline
      $g_\Gamma$ & $ap_z=0$ & $ap_z=2\pi/L$ & $ap_z=8\pi/L$ & $ap_z^{\text{phase}}=8\pi/L$ \\
      \hline
      $g_{A_z}^{u-d}$ & 1.18(1) & 1.145(9) & 0.8(1) & 1.275(29) \\
      $\chi^2_r$ & 1.255 & 1.421 & 12.301 & 2.761 \\
      \hline\hline
    \end{tabular}
  \end{center}
  \caption{Renormalized isovector axial charge determined via $\gamma_3\gamma_5$ at rest and in boosted frames.\label{tab:axialZTable}}
\end{table}

Other potential systematic errors in computations of $g_{A_z}^{u-d}$ have long been explored, such as use of $\mathcal{O}\left(a\right)$-improved currents~\cite{Liang:2016fgy}. In that work, however, it was found use of an $\mathcal{O}\left(a\right)$-improved axial current only mildly improved the experiment-lattice discrepancy, bolstering the presumed preponderance of excited-state and finite-volume effects. These same authors explored the degree to which $g_{A_z}^{u-d}=g_{A_4}^{u-d}$ could be satisfied, just as we now explore based on Eq.~\ref{eq:axialDecomp}.

\begin{figure*}[tb]
  \centering
  \subfigure[]{\includegraphics[width=0.49\linewidth]{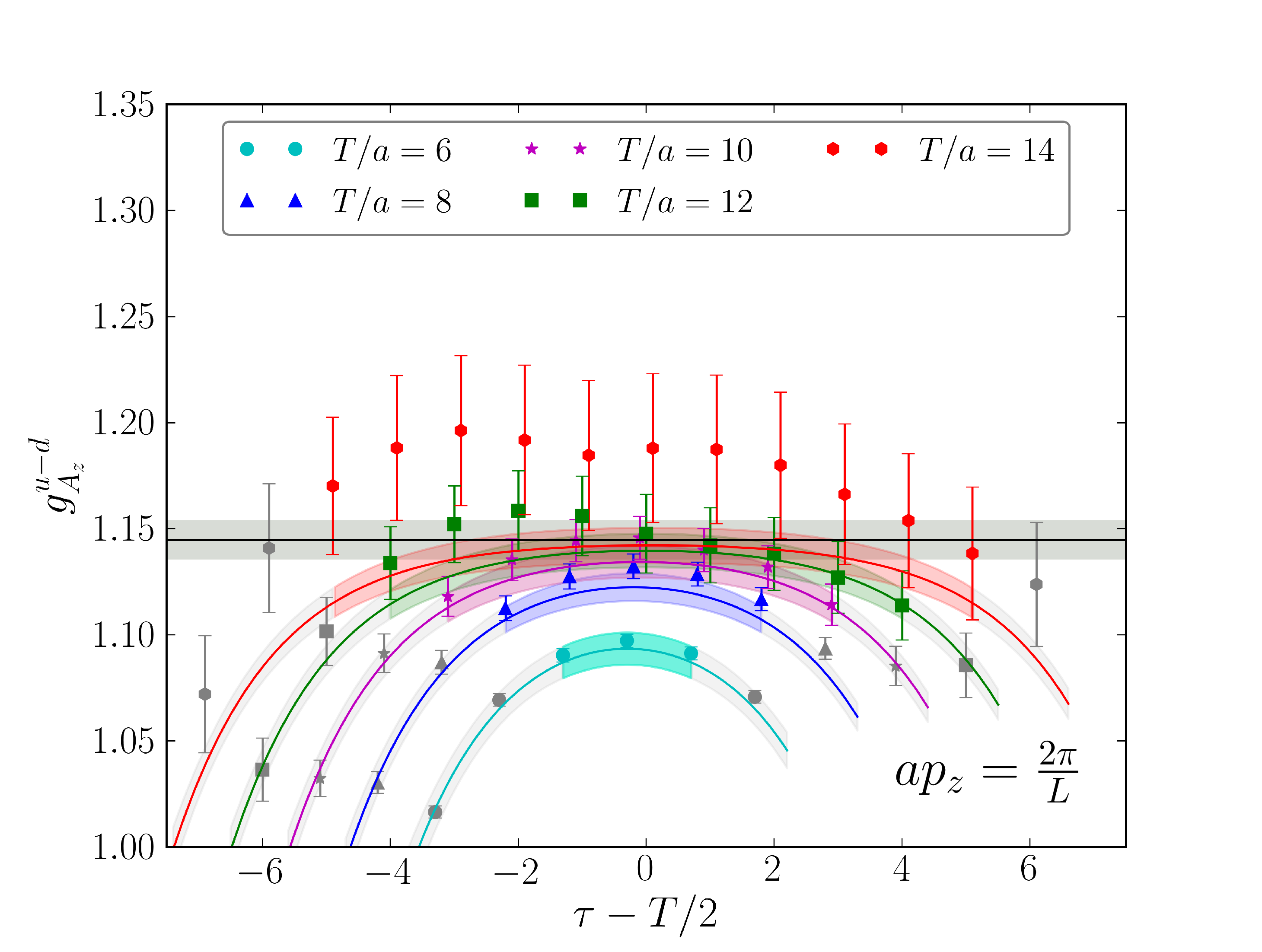}\label{fig:axialZ_p001_comp}}
  \subfigure[]{\includegraphics[width=0.49\linewidth]{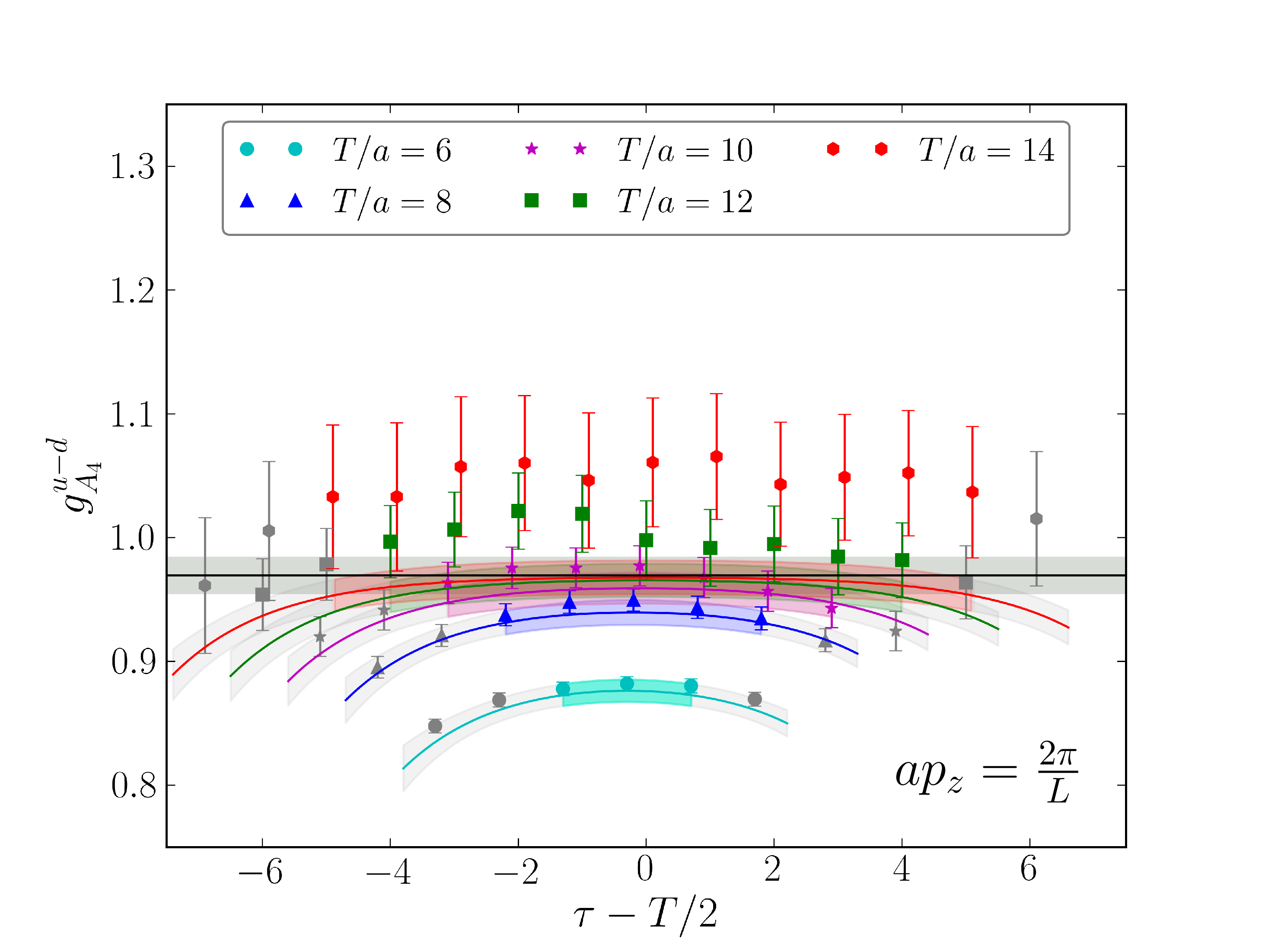}\label{fig:axial4_p001}}
  \subfigure[]{\includegraphics[width=0.49\linewidth]{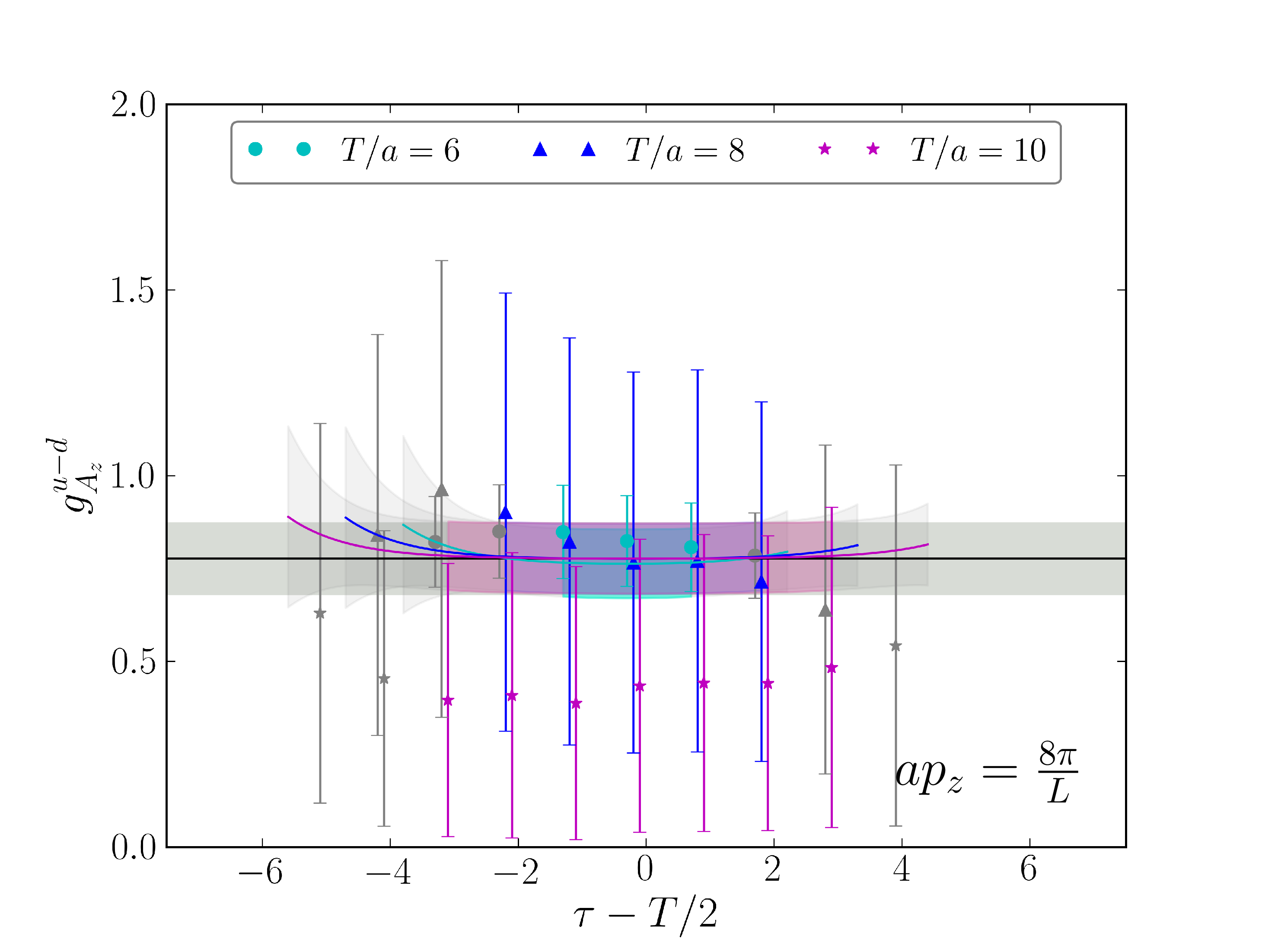}\label{fig:axialZ_p004_nophase_comp}}
  \subfigure[]{\includegraphics[width=0.49\linewidth]{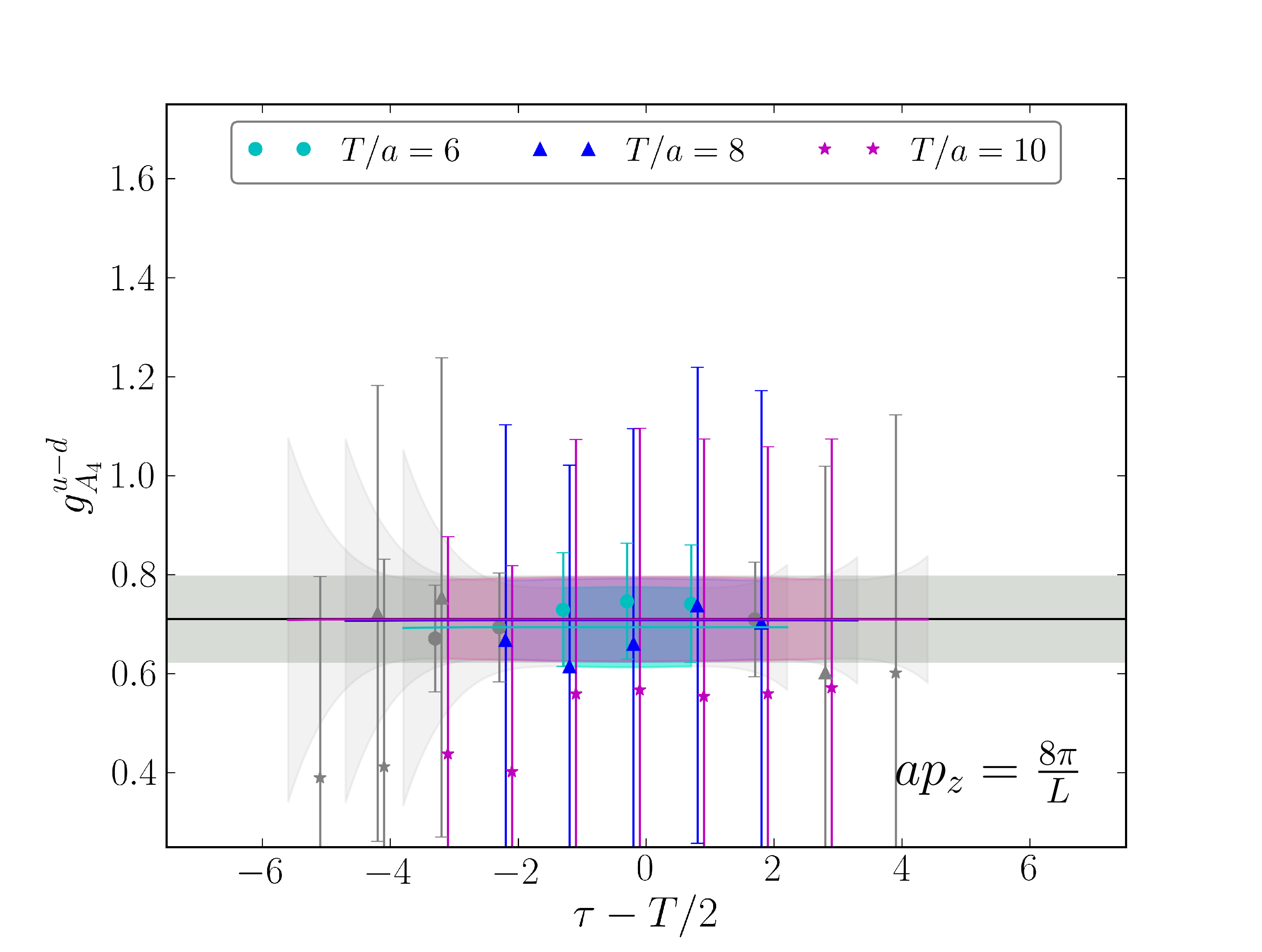}\label{fig:axial4_p004_nophase}}
  \subfigure[]{\includegraphics[width=0.49\linewidth]{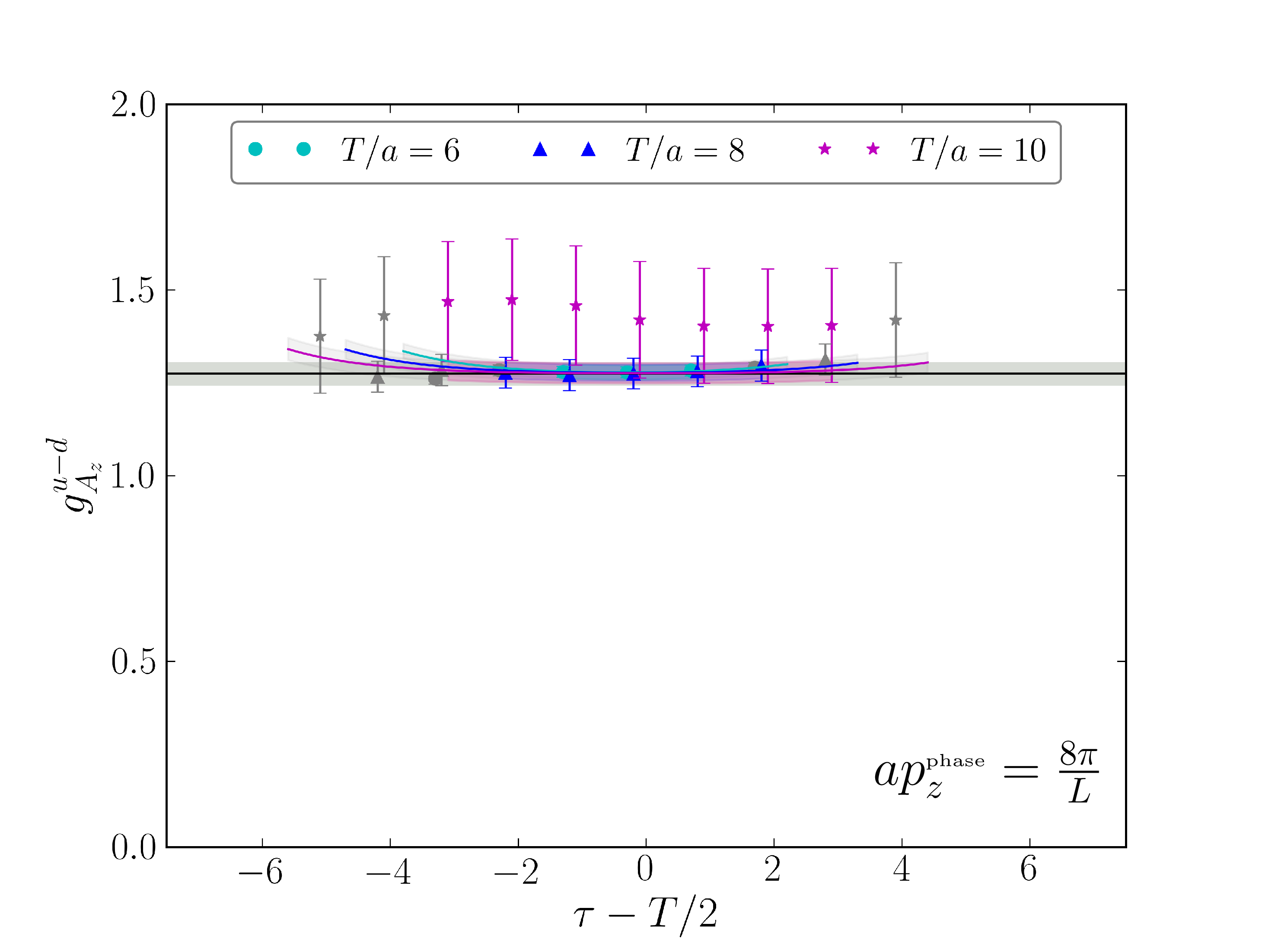}\label{fig:axialZ_p004_phase_comp}}
  \subfigure[]{\includegraphics[width=0.49\linewidth]{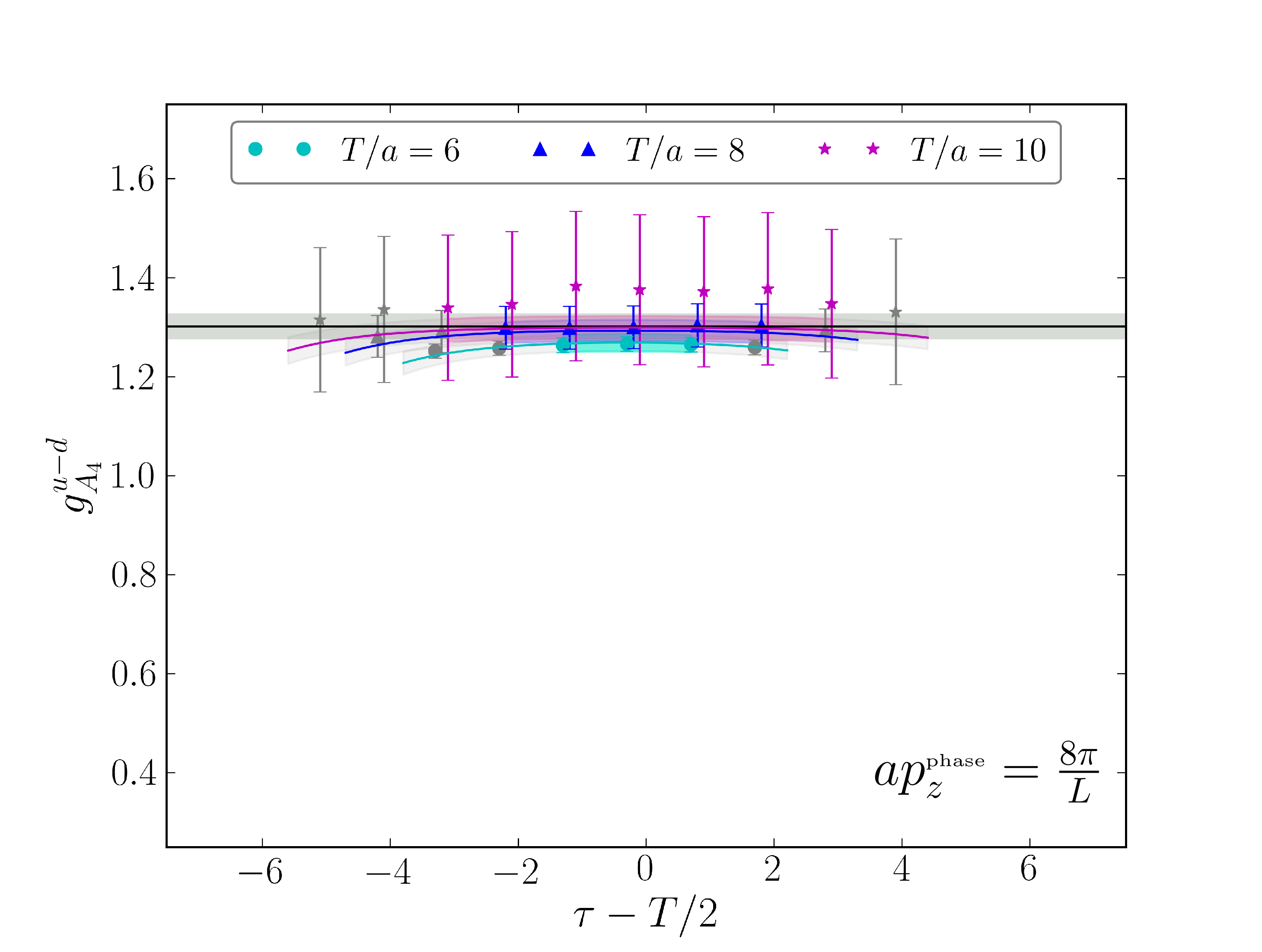}\label{fig:axial4_p004_phase}}  
  \caption{Extracted renormalized $R_{A_\mu}\left(T,\tau\right)$ and isovector axial charges using $\gamma_3\gamma_5$ (left) and $\gamma_4\gamma_5$ (right). External nucleon momentum according to {(a),(b)} $ap_z=\left(2\pi/L\right)$, {(c),(d)} $ap_z=4\left(2\pi/L\right)$ without phasing, {(e),(f)} $ap_z=4\left(2\pi/L\right)$ with two units of allowed lattice momentum applied to eigenvectors. Variationally improved operators were used within each momentum channel.\label{fig:axial_Z4_comp}}
\end{figure*}

Figure~\ref{fig:axial_Z4_comp} illustrates the $R_{\gamma_3\gamma_5}\left(T,\tau\right)$ and $R_{\gamma_4\gamma_5}\left(T,\tau\right)$ ratios isolated in the boosted frames we have considered. As with the scalar and vector charges, the lack of phasing at high-momentum degrades the two-point correlator such that the resulting matrix element signals contain essentially no information (Figs.~\ref{fig:axialZ_p004_nophase_comp} \&~\ref{fig:axial4_p004_nophase}). Compared to the rest frame, we observe $\sim3\%$ difference in $g_{A_z}^{u-d}$ when computed in the $ap_z=\left(2\pi/L\right)$ frame. This difference is indicative of $q^2\neq0$, despite $\vec{q}=0$, and hence mild radiative transitions with excited-states affecting this determination. Furthermore, we do observe a dramatic difference of $\sim15\%$ between the determination of $g_{A_z}^{u-d}$ and $g_{A_4}^{u-d}$ in the $ap_z=\left(2\pi/L\right)$ frame. This is unsurprising given the observation of $q^2\neq0$ in the moving $\gamma_3\gamma_5$ channel, all but ensuring the outsized influence of $\widetilde{G}_P^{u-d}$~\cite{Ishikawa:2018rew,Alexandrou:2017hac}. The increased separation between each $R_{\gamma_4\gamma_5}\left(T,\tau\right)$ in Fig.~\ref{fig:axial4_p001} again points to this increased excited-state contamination. We do remark that despite the different vertical scales chosen in Figs.~\ref{fig:axialZ_p001_comp} \&~\ref{fig:axial4_p001}, the fitted energy gap $\Delta E$ is consistent within error. The momentum smeared $ap_z=4\left(2\pi/L\right)$ charges (Figs.~\ref{fig:axialZ_p004_phase_comp} \&~\ref{fig:axial4_p004_phase}) again exhibit improved statistical quality, yet superficially appear to agree with each other and oddly with experiment. Each determination does not however seem to indicate a plateau in $R_\Gamma\left(T,\tau\right)$ has been found, especially in light of the noisy $R_\Gamma\left(T=10,\tau\right)$ determinations. Moreover, the $R_{\gamma_3\gamma_5}\left(T,\tau\right)$ and $R_{\gamma_4\gamma_5}\left(T,\tau\right)$ ratios are clearly trending away from each other within the illustrated data, and suggests the extracted charges in this phased frame would indeed be distinct were calculations performed with improved statistics and, especially, finer $T/a$. The results for our simultaneous fits for the $g_{A_z}^{u-d}$ and $g_{A_4}^{u-d}$ axial charges are presented in Tab.~\ref{tab:axialZTable} \&~\ref{tab:axial4Table}, respectively.

\begin{table}[th!]
  \begin{center}
    \setlength\abovecaptionskip{-1pt}
    \setlength{\belowcaptionskip}{-10pt}
    \begin{tabular}{c|c|c|c|c}\hline\hline
      $g_\Gamma$ & $ap_z=0$ & $ap_z=2\pi/L$ & $ap_z=8\pi/L$ & $ap_z^{\text{phase}}=8\pi/L$ \\
      \hline
      $g_{A_4}^{u-d}$ & -- & 0.970(14) & 0.71(9) & 1.302(24) \\
      $\chi^2_r$ & -- & 1.148 & 12.353 & 1.990 \\
      \hline\hline
    \end{tabular}
  \end{center}
  \caption{Renormalized isovector axial charges determined via $\gamma_4\gamma_5$ in boosted frames. By definition, $g_{A_4}^{u-d}=0$ at rest.\label{tab:axial4Table}}
\end{table}

\subsection{$g_T^{u-d}$}
\begin{figure*}[tb]
  \centering
  \subfigure[]{\includegraphics[width=0.49\linewidth]{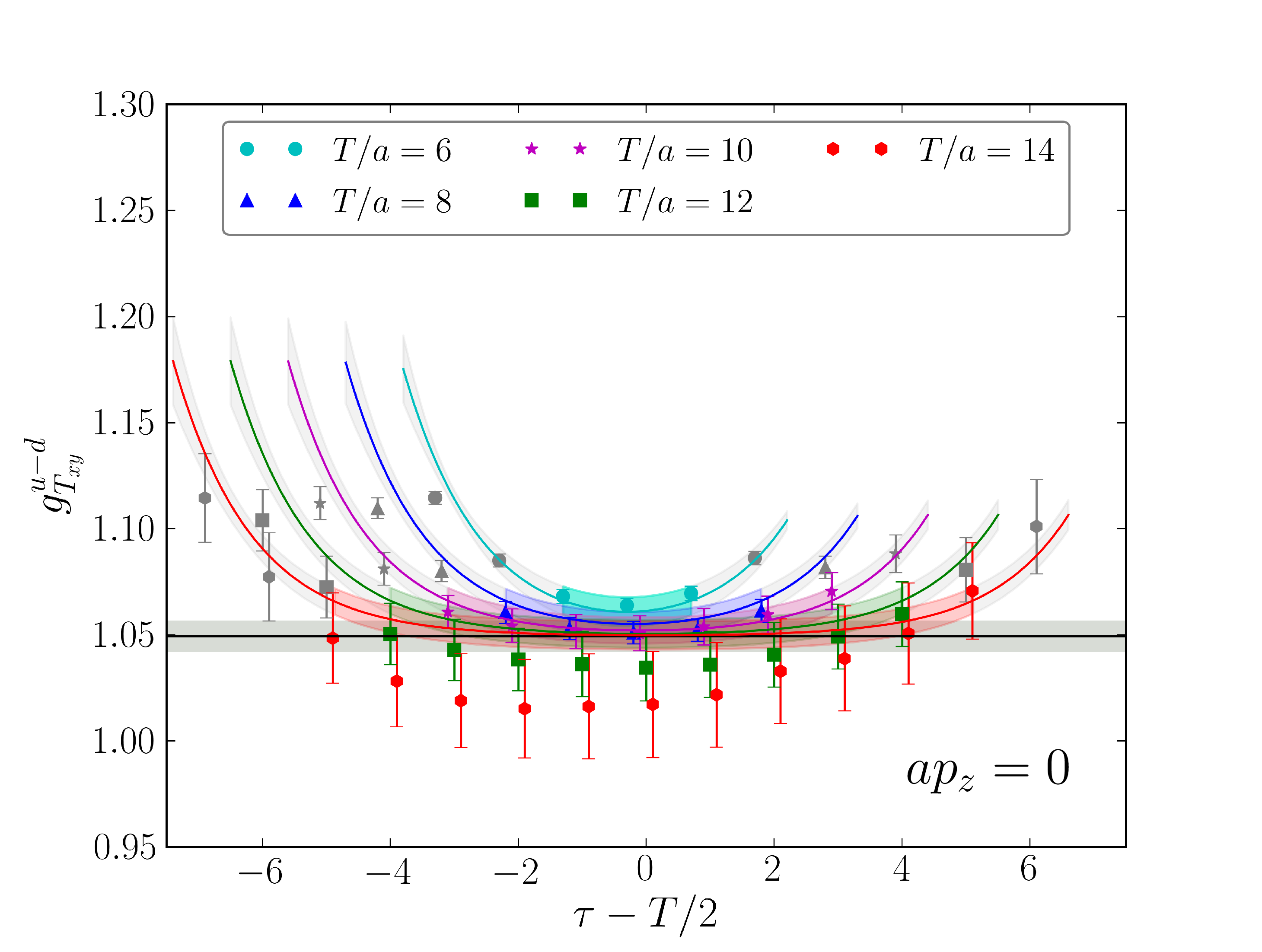}\label{fig:tensor12_p000}}
  \subfigure[]{\includegraphics[width=0.49\linewidth]{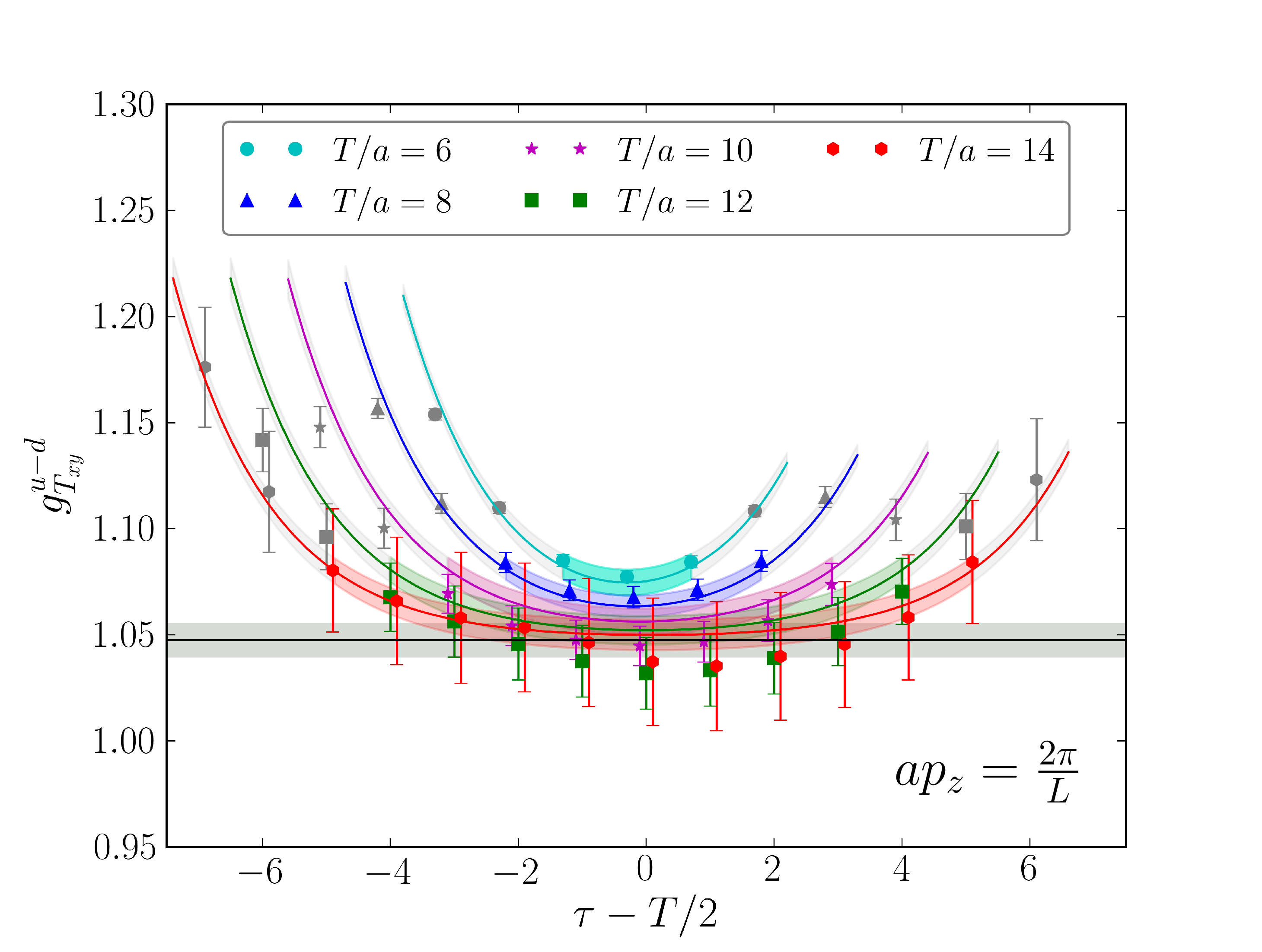}\label{fig:tensor12_p001}}
  \subfigure[]{\includegraphics[width=0.49\linewidth]{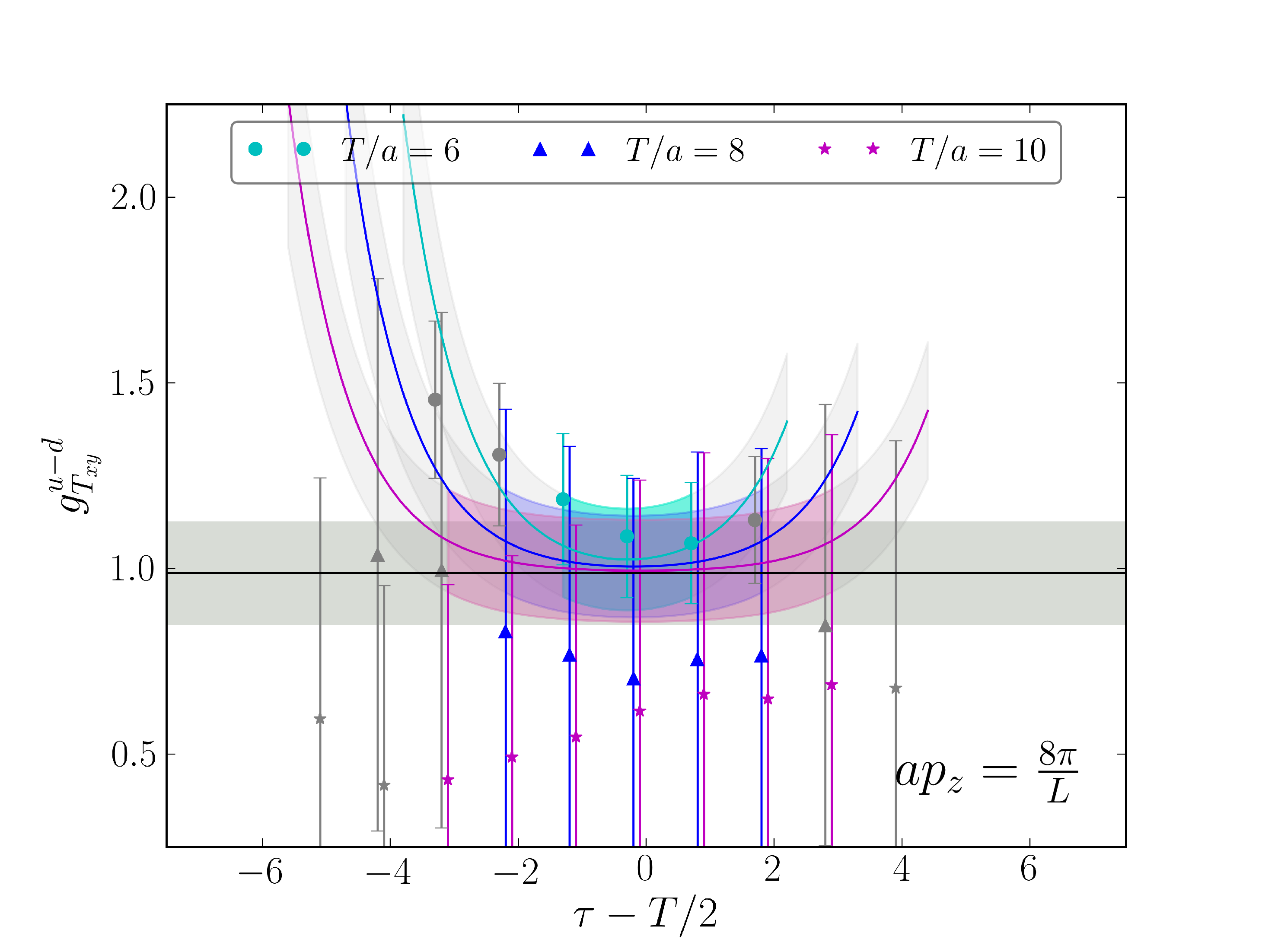}\label{fig:tensor12_p004_nophase}}
  \subfigure[]{\includegraphics[width=0.49\linewidth]{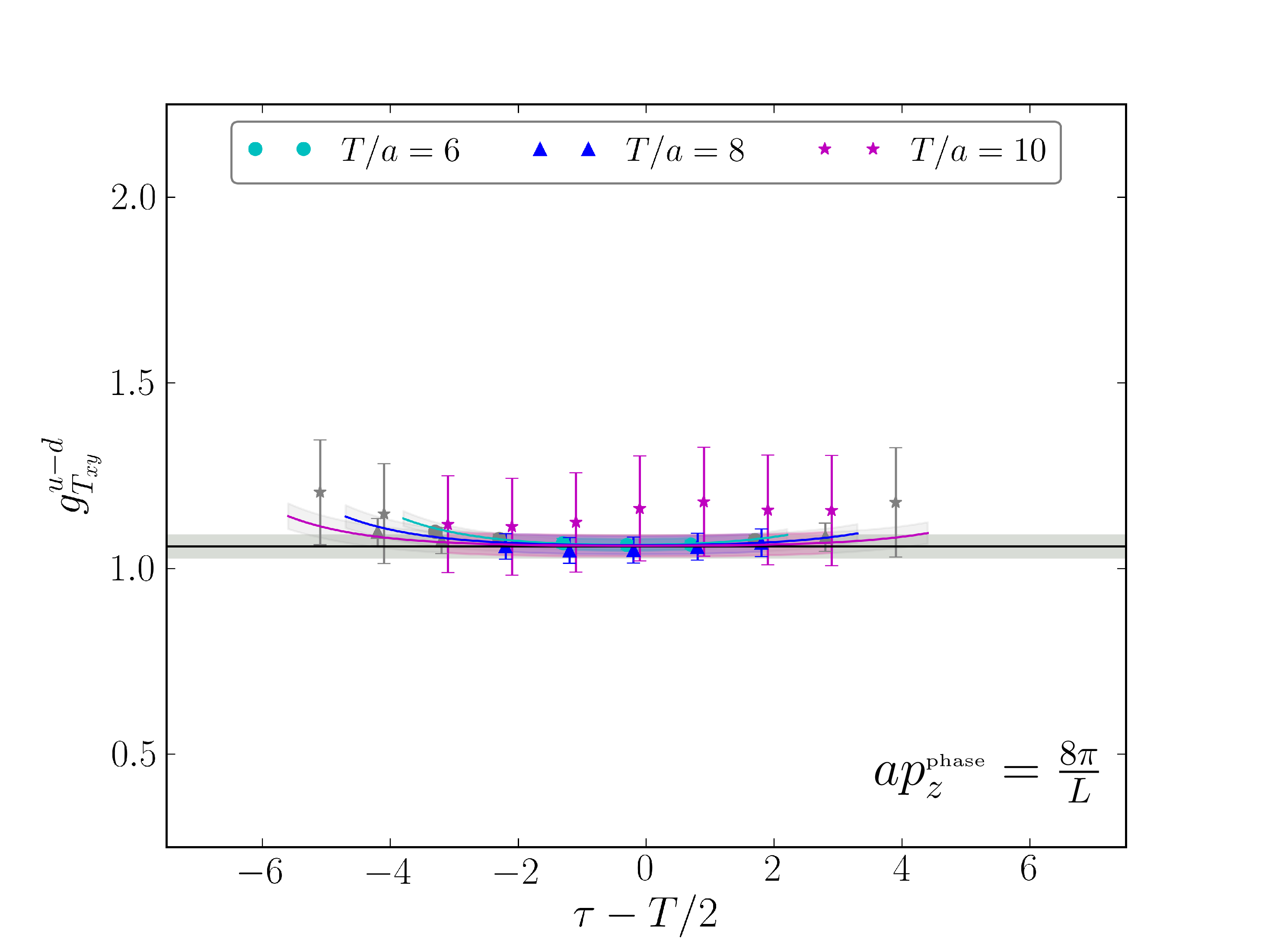}\label{fig:tensor12_p004_phase}}  
  \caption{Extracted renormalized $R_{T_{xy}}\left(T,\tau\right)$ and isovector tensor charge determined for (a) $ap_z=0$, (b) $ap_z=\left(2\pi/L\right)$, (c) $ap_z=4\left(2\pi/L\right)$ without phasing, and (d) $ap_z=4\left(2\pi/L\right)$ with two units of allowed lattice momentum applied to eigenvectors. Variationally improved operators were used within each momentum channel.\label{fig:tensor_comp}}
\end{figure*}

The isovector tensor current within nucleon states induces the following form factor decomposition:
\begin{align*}
  &\bra{N}T_{\mu\nu}\ket{N}=\overline{u}_N\left(p_f\right)\left[i\sigma_{\mu\nu}A_{10}^{u-d}\left(q^2\right)+\right. \\
    &\left.\quad\quad\quad\frac{\left[\gamma_\mu,q_\nu\right]}{2M_N}B_{10}^{u-d}\left(q^2\right)+\right. \\
    &\left.\quad\frac{\left[P_\mu,q_\nu\right]}{2M_N^2}\widetilde{A}_{10}^{u-d}\left(q^2\right)\right]u_N\left(p_i\right),
\end{align*}
where $T_{\mu\nu}=\overline{q}i\sigma_{\mu\nu}\tfrac{\tau^3}{2}q$ and $P=p_f+p_i$. At rest only the $T_{12}=\overline{q}\sigma_{12}\frac{\tau^3}{2}q$ matrix element is non-vanishing; apart from kinematic factors, this particular tensor current continues to be non-vanishing within the nucleon in motion. This has the fortunate consequence that all form factors outside the desired $A_{10}^{u-d}\left(q^2\right)$, where $g_{T_{xy}}^{u-d}\equiv A_{10}^{u-d}\left(0\right)$, do not contribute to the matrix element signal. We indeed find $g_{T_{xy}}^{u-d}$ determined in each momentum frame to be statistically consistent across the boosts considered, and in the case of $ap_z=\{0,1\}\times\left(2\pi/L\right)$ the charge is especially well determined and in fantastic mutual agreement (see Figs.~\ref{fig:tensor12_p000} \&~\ref{fig:tensor12_p001}). The lack of a clean signal for $g_{T_{xy}}^{u-d}$ for $ap_z=4\left(2\pi/L\right)$ (Fig.~\ref{fig:tensor12_p004_nophase}) is by now expected, and underscores the need for phasing at high-momentum (Fig.~\ref{fig:tensor12_p004_phase}). We note the slightly larger, though no less consistent, value found for $g_{T_{xy}}^{u-d}$ in the phased $ap_z=4\left(2\pi/L\right)$ frame appears to be a result of the noisy $T/a=10$ data. We anticipate future calculations with improved statistics will help to bring down this value. The ratios $R_{T_{xy}}\left(T,\tau\right)$ and simultaneous fit results are compared in Fig.~\ref{fig:tensor_comp} and the extracted tensor charges are gathered in Tab.~\ref{tab:tensorTable}. In the interest of completeness, we note our best determined $g_{T_{xy}}^{u-d}$ is $\sim8\%$ larger than $g_{T_{xy}}^{u-d}=0.973(36)$ determined in~\cite{Yoon:2016jzj}, yet several times more precise.

\begin{table}[tb]
  \begin{center}
    \setlength\abovecaptionskip{-1pt}
    \setlength{\belowcaptionskip}{-10pt}
    \begin{tabular}{c|c|c|c|c}\hline\hline
      $g_\Gamma$ & $ap_z=0$ & $ap_z=2\pi/L$ & $ap_z=8\pi/L$ & $ap_z^{\text{phase}}=8\pi/L$ \\
      \hline
      $g_{T_{xy}}^{u-d}$ & 1.049(7) & 1.048(8) & 0.99(14) & 1.06(3) \\
      $\chi^2_r$ & 1.267 & 1.064 & 12.603 & 1.999 \\
      \hline\hline
    \end{tabular}
  \end{center}
  \caption{Renormalized isovector tensor charges determined via $T_{12}$ at rest and in boosted frames.\label{tab:tensorTable}}
\end{table}

Summarizing, repeated calculations on lattice ensembles of varying lattice spacings and fixed physical volumes are necessary to rigorously pin down the size of discretization effects on these results. A dedicated study of contributing form factors is underway and will further facilitate the conclusions herein. Of course a further source of discrepancy of all computed charges are finite-volume effects. One would expect finite-volume effects to be minor for these charges, given that the $\Mlight$ ensemble is characterized by $m_\pi L\simeq4.24$. Nonetheless, calculations at different physical volumes required to confirm this expectation are planned.

\section{Conclusions\label{sec:Conclude}}
We have expounded upon the seminal Gaussian momentum smearing scheme developed by Bali et al., demonstrating momentum space overlaps of distilled interpolators can likewise be improved by introducing appropriate spatial phase factors onto eigenvectors of the gauge-covariant Laplacian. We elected to introduce phases onto a pre-computed eigenvector basis, rather than rotating the underlying gauge transporters. Consequently, the introduced phase factors were limited to allowed lattice momenta in the numerical investigations herein. Regardless of when the phase factors are introduced, all components forming the scaffolding of a distillation-smeared correlation function (e.g. elementals and perambulators) must be recomputed. This motivated our choice to smear pre-computed eigenvectors.

We established the efficacy of this approach by isolating the ground-state nucleon dispersion relation using a standard eigenvector basis, and two modified bases; modified with one and two units of allowed lattice momenta, respectively. Despite variational optimization of unmodified interpolators within the $J^\lambda=\tfrac{1}{2}^{\lambda=\pm1/2}$ channel, the nucleon dispersion relation was only meaningfully satisfied up to $\simeq1.75\text{ GeV}$. Variational analyses within the phase modified distillation spaces yielded agreement with the nucleon dispersion relation in excess of $3\text{ GeV}$.

The determination of several renormalized isovector charges of the nucleon was used as further evidence for the utility of merging distillation with momentum smearing. Matrix elements at rest and for $ap_z=\left(2\pi/L\right)$ were computed without phasing. These were then compared to identical matrix elements computed in a boosted frame ($ap_z=4\left(2\pi/L\right)$) with and without momentum phases. Our aim was to demonstrate consistency between charges computed in different (forward) frames. This is an especially nuanced venture, as numerous form factors begin to compete as the momentum frame is varied. Furthermore, the momentum smearing procedure certainly improves overlap onto unwanted single- and multi-particle excited states. A proper treatment of this consistency requires dedicated calculations of nucleon form factors at several lattice spacings/volumes, and pion masses. These encouraging results nevertheless establish the feasibility of future calculational paradigms requiring distillation at high-momenta. Our attention is now turned to such studies.

\subsection*{Acknowledgments}
We thank members of the \textit{HadStruc} Collaboration for invaluable discussions and scrutiny. We acknowledge the facilities of the USQCD Collaboration used for this research in part, which are funded by the Office of Science of the U.S. Department of Energy. This work used the Extreme Science and Engineering Discovery Environment (XSEDE), which is supported by the National Science Foundation under grant number ACI-1548562~\cite{tacc}. We further acknowledge the Texas Advanced Computing Center (TACC) at the University of Texas at Austin for HPC resources on {\tt Frontera} that have contributed greatly to the results in this work. We gratefully acknowledge computing cycles provided by facilities at William and Mary, which were provided by contributions from the National Science Foundation (MRI grant PHY-1626177), and the Commonwealth of Virginia Equipment Trust Fund. The authors acknowledge William and Mary Research Computing for providing computational resources and/or technical support that have contributed to the results reported within this paper. Calculations were performed using the {\tt Chroma}~\cite{Edwards:2004sx}, QUDA ~\cite{Clark:2009wm,Babich:2010mu}, QDP-JIT~\cite{Winter:2014dka}  and {\tt QPhiX}~\cite{ISC13Phi,qphix} software libraries  which were developed with support from the U.S. Department of Energy, Office of Science, Office of Advanced Scientific Computing Research and Office of Nuclear Physics, Scientific Discovery through Advanced Computing (SciDAC) program.
This research was also supported by the Exascale Computing Project (17-SC-20-SC), a collaborative effort of the U.S. Department of Energy Office of Science and the National Nuclear Security Administration.
This material is based upon work supported by the U.S. Department of
Energy, Office of Science, Office of Nuclear Physics under contract
DE-AC05-06OR23177. C.E. is supported in part by the U.S. Department of Energy
under Contract No. DEFG02-04ER41302, a Department of Energy Office of Science
Graduate Student Research fellowships, through the U.S. Department of Energy,
Office of Science, Office of Workforce Development for Teachers and Scientists,
Office of Science Graduate Student Research (SCGSR) program, and a Jefferson Science Associates graduate fellowship. The SCGSR program is administered by the Oak Ridge Institute for Science and Education (ORISE) for the DOE. ORISE is managed by ORAU under contract number DE-SC0014664.
 KO was supported in part by U.S. DOE grant  No. DE-FG02-04ER41302 and in part by the Center for Nuclear Femtography grants  C2-2020-FEMT-006, C2019-FEMT-002-05.

\clearpage
\bibliographystyle{/home/colin/texmf/tex/latex/commonstuff/bibtex/bst/revtex/apsrev4-1}
\bibliography{momsmear_srcs}
\end{document}